\documentclass[10pt]{article}

\usepackage[latin1]{inputenc} 
\usepackage[T1]{fontenc} 
\usepackage[english]{babel}

\usepackage{amsmath,amssymb}
\usepackage{graphicx}
\usepackage{subfigure}

\newcommand{\be}{\begin{equation}}
\newcommand{\ee}{\end{equation}}
\newcommand{\bea}{\begin{eqnarray}}
\newcommand{\eea}{\end{eqnarray}}

\def \la{\label}
\def\l{\ldots}
\def\({\left (}
\def\){\right )}
\def\]{\right]}
\def\[{\left[}
\def\<{\left <}
\def\>{\right>}
\newcommand{\cO}{\mathcal{O}}

\newcommand{\br}{\mathbf{r}}
\newcommand{\f}{\mathbf{f}}
\newcommand{\boe}{\mathbf{e}}
\newcommand{\bk}{\mathbf{k}}
\newcommand{\bA}{\mathbf{A}}
\newcommand{\bp}{\mathbf{p}}
\newcommand{\bq}{\mathbf{q}}
\newcommand{\bxi}{\pmb{\xi}}
\newcommand{\balpha}{\pmb{\alpha}}
\newcommand{\cW}{\mathcal{W}}
\newcommand{\ba}{\mathbf{a}}
\newcommand{\bR}{\mathbf{R}}
\newcommand{\e}{e} 
\newcommand{\Tr}{{\rm Tr}}

\newcommand{\aB}{a_{\scriptscriptstyle{B}}} 
\newcommand{\afs}{\alpha_\text{\mdseries fs}}
\newcommand{\bkhat}{\widehat{\mathbf{k}}}
\newcommand{\deltaT}{\delta^{\perp}}
\newcommand{\kB}{k_{\scriptscriptstyle {B}}}
\newcommand{\lat}{\lambda_\text{\mdseries at}}
\newcommand{\lcut}{\lambda_\text{\mdseries cut}} 
\newcommand{\lph}{\lambda_\text{\mdseries ph}}
\newcommand{\lscreen}{\lambda_\text{\mdseries screen}}
\newcommand{\Q}{{\cal Q}}
\newcommand{\R}{{\mathfrak{R}}}
\newcommand{\Zat}{Z_\text{\mdseries at}}

\newcommand{\moni}{{\rm i}}

\title{\Large{ \textbf{Atom-wall dispersive forces: a microscopic approach }}}
\author{
F. Cornu\\ Laboratoire de Physique Th\'eorique,  UMR 8627 du CNRS\\
Universit\'e Paris-Sud, B\^at. 210\\ F-91405 Orsay, France
\\ \vspace{3mm}\\
Ph. A. Martin\\
Institute of Theoretical Physics\\ Swiss Federal Institute for Technology
Lausanne\\ CH-1015, Lausanne EPFL, Switzerland}
\date{ \today}

\begin{document}
\maketitle

\begin{abstract}

We present a study of  atom-wall interactions in non-relativistic quantum electrodynamics by functional integral methods. 
The Feynman-Kac path integral representation is generalized to the case when the particle interacts with a radiation field, providing an additional effective potential that contains all the interactions induced by the field. We show how one can retrieve the standard van der Waals, Casimir-Polder and classical Lifshiftz forces in this formalism for an atom in its ground state. Moreover, when electrostatic interactions are screened in the medium, we find low temperature corrections that are not included in the Lifshitz theory of fluctuating forces and are opposite to them.

 \vskip 0.5cm 
{\it Corresponding author ~:} \\
CORNU Fran\c coise, \\
Fax: 33 1 69 15 82 87 \\
E-mail: Francoise.Cornu@u-psud.fr
\end{abstract}

\clearpage
\section{Introduction}

\subsection{Issue a stake}

Atom-wall interaction plays an important role in several physical, chemical and biological situations \cite{MahantyNinham1976,Parsegian2005} and has been the subject of active  measurement investigations during the last years (see \cite{AspectDalibard2002}  for a review and \cite{BezerraEtAl2008}  for references to more recent experimental results).
The basic model consists of a quantum atom in its ground state  at a distance $X$ from
a perfect metallic wall : the  atom interacts both with its wall mirror image and a source-free radiation field. 
In 1932 Lennard-Jones considered the case where interactions are purely electrostatic and predicted  a $1/X^3$ 
atom-wall interaction \cite{Lennard-Jones1932}; the latter is  similar to  the van der Waals interaction between two atoms in their  groundstates, first estimated by London in the framework of quantum mechanics and  classical electrodynamics \cite{London1930}.

In 1948 Casimir and Polder  adressed the case where the electromagnetic radiation field is quantized and they calculated the atom-wall interaction  at the second order of the ground-state perturbation theory  \cite{CasimirPolder1948}.  For $X$ small (a few nanometers) they showed that the atom-wall potential is the   electrostatic dipolar  $1/X^3$ van der Waals-like interaction, with an amplitude given by  the ground-state  position fluctuations inside the isolated atom. At larger distance (a few micrometers) they found a cross-over to a longer-ranged $1/X^4$ potential, with an amplitude determined by the vacuum fluctuations of the  source-free radiation field  confined by the metallic wall. When calculations are performed in the Coulomb gauge of electrodymanics, the interpretation in terms of the typical  atomic-line wavelength $\lat$ is as follows. At  distances $X\ll \lat$  the Coulomb image interaction dominates the atom-radiation interaction (of relative order $X/\lat$). However at  distances $X\gg \lat$ the atom-radiation interaction becomes of the same order as the Coulomb interaction. In fact its leading term becomes opposite to the Coulomb image interaction, while its subleading term  (of relative order $\lat/X$) does not involve the instantaneous interactions artificially introduced by the Coulomb gauge : the $1/X^4$ Casimir-Polder tail proves to be a retardation effect.

 For an atom  prepared in an excited state and which  interacts with the photon field in its vacuum state,
 the atom-wall interaction  has been investigated at the second order in  the atom-field coupling in
 Ref.\cite{MeschedeEtAl1990}.  The authors used the microscopic approach of Ref.\cite{DalibardEtAl1982,DalibardEtAl1984} (see also \cite{Cohen-TannoudjiEtAl1992}) originally introduced to disentangle the two phenomena that come out in the source-field picture: on one hand  the ``{self-reaction}'' effect, due to the interaction between the electron and its own field, on the other hand, the ``{vacuum fluctuation}'' effect, due to the interaction between the electron and the source-free part of the  quantized electromagnetic field. The short-distance interaction is again dominated by  the van der Waals-like interaction, which emerges as a purely self-reaction contribution. At large distances  the leading net atom-wall interaction becomes  an  oscillating $1/X$ decay : this tail  is similar to the interaction between  a classical oscillating electric dipole and its own reflected far field. However it arises only from lower-lying states, as in the spontaneous-emission rate,
 because it results from the combination of self-reaction and  vaccum-fluctuation contributions.
 For an atom in its ground-state, which cannot spontaneously radiate,   the oscillating $1/X$ tail proves to vanish and the  atom-wall interaction is reduced to the subleading contribution, namely the  Casimir-Polder potential which results from  the change in the spectrum of vacuum fluctuations due to the reflection of the source-free radiation field on the metallic boundary.
 
 In the case of an atom in equilibrium with a thermalized photon field, the atom-wall interaction  has been studied for a simplified model of the atom in Ref.\cite{MendesFarina2007} 
 by using the   microscopic approach of Ref.\cite{DalibardEtAl1982,DalibardEtAl1984} again. The atom in the model contains only two atomic levels, and the  limitations of this simplication have been pointed out in \cite{Barton1974}.
The  energy shift of the  excited state has an oscillating $1/X$ tail, but, after thermal equilibrium average, this contribution is exponentially damped at low temperature. Moreover, at distances larger than the length scale $\lph$ beyond which thermal fluctuation contributions  dominate over vacuum ones,
the energy-shift $X$-dependence is changed into the  $1/X^3$ classical Lifshitz potential \cite{Lifshitz1956} which arises only at non-zero temperature. Indeed, nowadays
the semi-macroscopic Lifshitz theory \cite{Lifshitz1956,LifshitzPitaevskii1980} (see also \cite{Parsegian2005})
is the commonly adopted theoretical framework for calculating forces induced by quantum and thermal fluctuations. A detailed application of Lifshitz theory to the atom-wall interaction can be found in the recent paper \cite{BezerraEtAl2008} (see references therein).

In the present work, we  investigate the  low-temperature  atom-wall interaction  for an atom in a dilute gas in equilibrium with the photon field  when the internal structure of the atom  is described from the full Hamiltonian of quantum electrodynamics with nonrelativistic matter. 

We revisit the original Casimir-Polder model with a method based on functional integration which is in principle not perturbative and not limited to zero temperature. The model consists of an Hydrogen atom with spinless electron of charge $e$ and mass $m$, and an infinitly massive proton. 
For a single atom in thermal equilibrium with the electromagnetic field at inverse temperature $\beta=(k_{B}T)^{-1}$, the atom-wall potential $\Phi(X,\beta)$  is defined as the excess free energy of the atom immersed in the photon field at a distance $X$ from the wall. The calculation of $\Phi(X,\beta)$ involves the trace over the photonic degrees of freedom of the quantum Gibbs factor.
The main tool is the functional representation of the thermal equilibrium weight of a quantum particle submitted to an external static potential by means of the Feynman-Kac-It\^{o} formula \cite{Simon1979, Roepstorff1994}. When the external field fluctuates, the integration over the fluctuations leads to a generalized  Feynman-Kac-It\^{o} formula with an additional effective potential. 
Using a bosonic functional integral representation of the thermalized photon field, we establish  such a generalized  Feynman-Kac-It\^{o} formula  for the electron in interaction with the quantum electromagnetic field in some region with general  boundary  conditions (see \eqref{ToutF} and \eqref{multimode}). 
In an homogeneous system where the radiation field extends over  the whole space, we recover the results of
\cite{BoustaniEtAl2006} for a classical field and \cite{BuenzliEtAl2007} for a quantum field.

\subsection{Finite-temperature effects}

As soon as the temperature is different fron zero, numerous effects
come into play and we have to investigate them separately.
Among them we distinguish
\begin{enumerate}
\item The spectral broadening and level shifts of the atom due to radiative interactions with the photons. 
\item The  thermal excitation of the atom and its possible ionization.
\end{enumerate}
All these effects are embedded in the effective potential $\Phi(X,\beta)$. In this paper we determine the dominant term of $\Phi(X,\beta)$ for $X$ large compared to various microscopic lengths in the system, while keeping only lowest order effects in the fine structure constant $\afs$. This will amount to neglect spectral broadening and level displacements 
in final calculations and deal with the usual energy levels $E_{i}, \; i=0, 1,\ldots$ of the bare Hydrogen atom. Moreover it also corresponds to work in the so-called dipolar approximation (as in the original Casimir-Polder paper), where the radiation-field spatial fluctuations inside the atome are neglected. The study of relativistic diamagnetic terms, that arise beyond the dipolar approximation, is postponed to a forthcoming paper \cite{CornuMartin}. 

Concerning the second point, one must observe that
the notion of atomic bound state makes sense only if the thermal energy is much less than the ionization threshold $|E_0|$, namely $k_{B}T\ll |E_0|$. We shall therefore consider the low-temperature regime characterized by
\begin{equation}
k_{B}T\ll E_1-E_0
\label{lowtemp}
\end{equation}
when the thermal energy is insufficient on the average to excite the atom from its ground state. 
More precisely, we shall neglect exponentially vanishing temperature corrections ${\cal O}(e^{-\beta(E_{1}-E_{0})})
$ and retain only those that are algebraically growing with  $T$.

 At the microscopic level, even if the low-temperature condition \eqref{lowtemp} is satisfied, there is always a non-vanishing probability
to populate Rydberg states and to ionize the atom.  As a consequence a summation  on the atomic thermal weights is divergent. 

In reality envisaging a single thermalized atom does not make physical sense : one has to consider a  dilute gas of atoms at positive density. In addition to thermal photons, at any non-zero temperature the atoms in the gas  are also in equilibrium with an ionized fraction of electrons and nuclei that provide collective screening mechanisms on the length scale  $\lscreen$. 
For a purely Coulombic quantum gaz at low density and low temperature, it has been shown that collective screening has two main consequences : 
\begin{description}
\item[(i)] a natural regularization of atomic traces that makes them finite,
\item[(ii)] the introduction of an appropriate screened Coulomb potential. 
\end{description}
These results  come from the elaborate discussion of the subtle interplay between quantum mechanical binding, ionization and screening presented in \cite{ABCM2003}  as well as from the analytical investigation  of the  effective interactions in a partially-ionized dilute gas at  low temperature in \cite{ACM2007} or the effective charge-wall interaction in an ionized dilute gas \cite{AquaCornu2004I}. 

Although we presume that the same studies extend to the present electromagnetic system with the same 
conclusions, we shall not enter here in 
the full many-body problem in order to keep the presentation as simple as possible.
For the finiteness of traces (point (i)), we rather introduce a formal spatial cut-off when calculating the traces of atomic observables by limiting the spatial integration over the electron position $\br$ to a ball of radius $R_{0}$ centered around the  proton position $\bR$, $\int d\br\rightarrow\int d\br_{|\br-\bR|\leq R_{0}}$. This cut-off, of the order of $R_{0}=\rho^{-1/3}$ ($\rho$ the  density of the screening gas), can be interpreted as delimiting the effective available configurational space for an atom when the density is different from zero. Equivalently, when working in the energy representation, we shall cut the level sums at some maximal energy $E_\text{\mdseries max}$.  
When $X\ll \lscreen$, we calculate $\Phi(X,\beta)$ in the low-temperature regime (\ref{lowtemp}) using the bare Coulomb potential. It turns out that in this regime, results are independent of the cut-off $R_{0}$ which can eventually be removed. When
$X\gg \lscreen$ we procceed in the same way, but take collective screening into account by replacing the bare Coulomb potential between the atom and its image by the screened potential mentioned in point (ii) above.  This  screened potential, extensively studied in \cite{BMA2002}, does not follow the classical exponential Debye law, but it 
has an algebraically decaying tail. This tail, generated by the intrinsic quantum fluctuations of the particles \cite{BrydgesMartin1999}, is dominating when $X\gg\lscreen$.
It is of dipolar type and therefore should not be omitted 
in the calculation of the atom-wall interaction  \cite{AquaCornu2004I}. 

The $X$-behaviour of the atom-wall interaction is also determined by the  hierarchy of the length scales 
in  the  microscopic  model.
First, since the electron is non relativistic, one has to disregard high-energy photons that could generate high-energy processes such as pair creation, namely we consider only photons with $\hbar c k\ll mc^2$ ($k$ is the photon wave number and $c$ the speed of light). 
This introduces an ultraviolet wavenumber cut-off $k_{\text{cut}}=(\lcut )^{-1}= mc/\hbar$ on the field modes.
Next, the length scales associated to the atomic properties are the Bohr radius $\aB=\hbar^2/m e^2 $ and the atomic wavelength $\lat=\hbar c/(E_1-E_0)$ corresponding to the photonic transition between the ground state and the first excited atomic level. Finally $\lph= \beta \hbar c$ defines the thermalization length of the photon.  
Since $E_1-E_0=3|E_0|/4$ and $|E_0|=e^2/2\aB$, one has
$\lcut /\aB=\afs$ and $\aB/\lat=(3/8)\afs$ where
$\afs=e^2/\hbar c$ is the fine structure constant, so that
\begin{equation}
\lcut \ll \aB\ll\lat
\label{scale1}
\end{equation}
Since $\lat/\lph=k_{B}T/(E_1-E_0)$, the condition (\ref{lowtemp}) implies also that  $ \lat\ll\lph$ and with (\ref{scale1}) the full hierarchy of length  scales reads
\begin{equation}
\lcut \ll \aB\ll\lat\ll\lph
\label{scale2}
\end{equation}
 However it will be always assumed that the wall separation is larger than the atomic size, $X\gg \aB$ (the model does not make sense for $X\sim \aB$ since the wall is described at the macroscopic level).

\subsection{Scheme of the paper and statment of results}

The paper is organized as follows.
First, in section 2  the generalized  Feynman-Kac-It\^{o} formula with the additional effective potential  that embodies the interactions induced by the quantized field is derived for general boundary conditions. The quantum photonic oscillators are represented by the Gaussian oscillator stochastic processes. In this representation, the electron appears as submited to an external classical
random field and the standard Feynman-Kac-It\^{o} formula applies. Then the field degrees of freedom can be integrated out by means of a simple Gaussian integral.
The procedure is well known and can be found in \cite{Roepstorff1994} and \cite{Spohn2004} page 187.

In section 3, we specify the above formula to the atom-wall model
by introducing the mirror charges and fixing
the field boundary conditions at the metallic wall. The  expression of the  atom-wall potential $\Phi(X,\beta)$ in terms of  path integral is given in \eqref{Phifinal}.
At this point, the  large-distance analysis is not perturbative with respect to the dimensionless coupling constant $\afs$. For instance,
formulae (\ref{defPhiasa}) and (\ref{defPhiasb}) contain several
higher-order effects in $\afs$ such as diamagnetic
polarization terms and thermal displacement and broadening of spectral lines of the atom in the photon field.

In section 4 we consider a density  regime where $\lscreen$ is larger than all length scales in the hierarchy \eqref{scale2} and fix the temperature in the low-temperature regime (\ref{lowtemp}) where exponentially-small temperature corrections are negligible compared to the algebraic ones. 
We show how our formalism allows one to recover the van der Waals \eqref{Phicas}, Casimir-Polder \eqref{retarded2}
and classical Lifshitz \eqref{classical2} potentials  in the respective ranges $X\ll\lat$, $\lat\ll X\ll\lph$ and $\lph\ll X\ll \lscreen$.  For this aim we neglect the above mentioned finer relativistic effects by making the dipolar approximation 
and by switching off the atom-field coupling in the bulk. Then the functional integrals can be expressed in terms of thermal averages of atomic observables.

In section 5 first we analyze the effect of screening when $X\gg \lscreen$ by introducing the screened Coulomb potential.
Various cases can occur according to the possible value of $\lscreen$ compared to the other lengths. When $\lscreen\gg \lph$
we show that the classical Lifshitz potential is exactly canceled by the  thermal screening correction linear in $T$. Only diamagnetic terms beyond the dipolar approximation do survive in the leading tail of the atom-wall interaction (these terms will be presented in \cite{CornuMartin} ).
For densities such that either  $\lscreen\ll \lph$ or even $\lscreen\ll\lat$, the same thermal screening correction linear in $T$ has to be added to the Casimir-Polder or van der Waals  potentials whenever    $\lscreen\ll X\ll \lph$ or $\lscreen\ll X \ll\lat$  respectively. More comments are offered in the concluding remarks.

\section{A generalization of the  Feynmann-Kac formula to retarded interactions}

We consider an electron submitted to an external potential $V(\br)$ and in interaction with the quantum electromagnetic field. With $\bq,\bp$, $[q^\mu,p^{\nu}]=\moni\hbar\delta_{\mu\nu},\;\mu,\nu=1,2,3$,
the canonical quantum variables of the electron,
the total Hamiltonian written in the Coulomb gauge is 
\begin{align}
H=\frac{1}{2m}\(\bp-\frac{e}{c}\bA(\bq)\)^2 +V(\bq)+ H_\text{rad},\quad\quad\nabla\cdot \bA=0
\label{3.1}
\end{align}
The field is enclosed in a region $\Lambda$ and obeys appropriate boundary conditions at the border of $\Lambda$. These boundary conditions, which do not need to be specified at this point, define a complete set of (real) orthogonal and divergence free field eigenmodes 
$\f_{\gamma}(\br)
, \int_{\Lambda} d \br\f_{\gamma}(\br)\f_{\gamma'}(\br)=0$ if $ \gamma\neq\gamma', \;\nabla\cdot\f_{\gamma}(\br)=0$. The eigenmode   expansions of the vector potential and of the free photon Hamiltonian read
\begin{align}
\bA(\br)&=\sum_{\gamma}(a_{\gamma}+a^\dag_{\gamma})\f_{\gamma}(\br)
\label{3.1b}\\
H_\text{rad}&=\sum_{\gamma}\epsilon_{\gamma}a_{\gamma}^\dag a_{\gamma}
\label{3.1c}
\end{align}
where $a_{\gamma}^\dag,a_{\gamma}$ with $[a_{\gamma},a_{\gamma'}^\dag]=\delta_{\gamma\gamma'}$, are the photon creation and annihilation
operators in the mode $\gamma$ with corresponding energy $\epsilon_{\gamma}$. Normalization factors entering into the definition of the vector potential are included in $\f_{\gamma}$.
\bigskip

The electron and the photons are supposed to be in thermal equilibrium at a common inverse temperature $\beta=(k_{B}T)^{-1}$ ($k_{B}$   the Boltzmann constant), and we are interested in a functional integral representation 
of the effective electronic Gibbs weight when the field degrees of freedom have been traced out
\be
\label{defAverageRad}
 \langle e^{-\beta H}\rangle_\text{rad}
 \equiv\frac{1}{Z_{\text{rad}}}\Tr_\text{rad} e^{-\beta H}
\ee
where $\Tr_\text{rad}$ denotes the partial trace on the Fock space of photon states and
$Z_{\text{rad}}=\Tr_{\text{rad}}\e^{-\beta H_\text{rad}}$ is the partition function of the free photon field.
To this aim, it is convenient to first introduce a functional integral representation of the action of the field variables. 

\subsection{Path integral representation for photonic modes : the oscillator process}

For simplicity we deal first with a single mode $\f(\br)$
of the field having energy $\epsilon$ (dropping the mode index $\gamma$). We introduce the dimensionless canonical variables $(Q,P)$ of the mode, setting $a+a^\dag=\sqrt{2 \beta\epsilon}Q,\; \moni (a^\dag-a)=\sqrt{2/\beta\epsilon}P$ so that $[Q,P]=\moni$. 
The dimensionless free Hamiltonian of this mode
becomes
\begin{equation}
\beta H_{0}=(\beta\epsilon) a^\dag a=\frac{1}{2}(P^2+(\beta\epsilon)^2Q^2-\beta\epsilon) 
\label{3.2}
\end{equation} 
and the contribution of this mode to the vector potential is
$\bA(\br)=\sqrt{2 \beta\epsilon}Q\;\f(\br)$.

The trace over  the states of  one  photon mode can be performed as an integral on the partial configurational oscillator matrix elements $(R\vert e^{-\beta H}\vert R)$, where $\vert R)$ is an eigenstate of the operator $Q$, so the  partial average \eqref{defAverageRad} reads 
\be
\label{Rintegral}
 \langle e^{-\beta H}\rangle_\text{rad}=\frac{1}
 {Z_{0}}\int dR  (R\vert e^{-\beta H}\vert R)
\ee
with $Z_{0}=\int dR  (R\vert e^{-\beta H_{0}}\vert R)$.
To calculate these matrix elements we apply the method of path integral for bosons developed in chapter 5 of \cite{Roepstorff1994}, see also  \cite{Spohn2004}. 
First, we split the total Hamiltonian as
\begin{align}
H=H_\text{el} +H_{0}
\label{3.7}
\end{align}
where the electronic part 
\begin{align}
H_\text{el}=\frac{1}{2m}\(\bp-\tfrac{e}{c}\sqrt{2 \beta\epsilon}Q\;\f(\bq)\)^2 +V(\bq)=H_\text{el}(Q)
\label{3.8}
\end{align}
depends both on the field operator $Q$ and the electronic operators. 
The Trotter product formula applied to the partial matrix element $(R|\e^{-\beta H}|R)$ reads
\begin{eqnarray}
&&(R|\e^{-\beta H}|R)=(R|\e^{-\beta H_\text{el} -\beta H_0}|R)
=\lim_{N\to\infty}(R|
[e^{-\beta H_\text{el}/N}
\e^{-\beta H_0/N}]^N|R)
\nonumber\\
&&
=\lim_{N\to\infty}\int d R_{N-1}...\int d R_{n}...\int d R_{1}\;
\e^{-\beta H_\text{el}(R)/N }(R|\e^{-\beta H_0/N }|R_{N-1})\cdots 
\nonumber\\&&
\times \e^{-\beta H_{\text{el}}(R_{n})/N }(R_{n}|\e^{- \beta H_0/N }|R_{n-1})\cdots
\e^{-\beta H_\text{el}(R_{1}) /N}(R_{1}|\e^{-\beta H_0/N }|R)
\label{Trotter}
\end{eqnarray}
A product of configurational matrix elements of the free oscillator 
\begin{align}
(R|\e^{-(s'-s_{N})\beta H_0}|R_{N})\cdots(R_{n}|\e^{-(s_{n}-s_{n-1})\beta H_0}|R_{n-1})\cdots(R_{1}|\e^{-(s_{1}-s)\beta H_0}|R)
\la{3.6}
\end{align}
defines the oscillator process. The product (\ref{3.6})
is interpreted as the joint probability (up to a normalization)
for a closed path $\R(s)$ starting in $R$ at time $s$ to be found in $R_{1}$ at $s_{1}\ldots,$ in $R_{n}$ at $s_{n}\ldots,$ and again in $R$ at time $s'$.  
The matrix element $(R'|\e^{-(s'-s)\beta H_0}|R)$ is given by the Mehler formula (see \cite{Roepstorff1994}  section 1.8.1) which is a Gaussian function of $R$ and $R'$, so that the process defined by the distributions \eqref{3.6} 
 is Gaussian.\footnote{The explicit form of Mehler formula will not be used here. } The corresponding 
Gaussian functional integral is 
 denoted by $\int_{\R(0)=R}^{\R(1)=R}D[\R]\cdots$.

Introducing the time dependent electronic Hamiltonian
\begin{equation}
\label{defHelRs}
H_\text{el}(\R(s))=\frac{1}{2m}\[\bp-\tfrac{e}{c}\sqrt{2 \beta\epsilon}\R(s)\;\f(\bq)\]^2 +V(\bq)
\end{equation}
associated with a given realization of the process, we see that the partial matrix element (\ref{Trotter}) can be represented by the functional integral
on oscillator paths $\R(s)$
\bea
\nonumber
(R\vert e^{-\beta H}\vert R)
=\int_{\R(0)=R}^{\R(1)=R}D[\R]{\cal T}\exp\left[-\beta\int_{0}^{1} ds H_\text{el}(\R(s))\right]
\eea
The need for the chronological ordering ${\cal T}$ in the propagator comes from the fact that 
 $H_{\text{el}}(\R(s))$ is still an operator depending on the canonical variables of the electron.
The partial average \eqref{Rintegral} eventually  reads 
\be
\langle e^{-\beta H}\rangle_\text{rad}
=\left\langle{\cal T}\exp\left[-\beta\int_{0}^{1} ds H_\text{el}(\R(s))\right]\right\rangle_{\text{rad}}
\label{3.10}
\ee
where $\left\langle\cdots\right\rangle_\text{rad}$ is to be evaluated
by the normalized integral
\begin{align}
\left\langle\cdots\right\rangle_\text{rad}=\frac{1} 
{Z_{0}}\int dR\int_{\R(0)=R}^{\R(1)=R}D[\R]\cdots
\label{3.11}
\end{align}
when field quantities are expressed as functionals of the paths
$\R(\cdot)$. (Then $Z_{0}$ defined in \eqref{Rintegral} also reads 
$Z_{0}=\int dR\int_{\R(0)=R}^{\R(1)=R}D[\R]$.)
The stationnary Gaussian process (\ref{3.11})  is entirely defined by its covariance $\left\langle \R(s)\R(s')\right\rangle_\text{rad}=\left\langle \R(|s-s'|)\R\right\rangle_\text{rad}
$. The latter is easily calculated when we express it in operator form
\begin{align}
\left\langle \R(s)\R(s')\right\rangle_\text{rad}
=\frac{1}{Z_0}
\Tr\left[\e^{-\beta H_0}{\cal T}(Q(s)Q(s'))\right]
\label{3.12}
\end{align}
where $Q(s)=\e^ {s \beta H_0}Q\e^{-s \beta H_0}$ is the imaginary time evolved operator.
Using the commutation relations
\begin{align}
\frac{d}{ds}Q(s)&=[\beta H_0,Q](s)=-\moni P(s)\nonumber\\
\frac{d^2}{ds^2}Q(s)&=[\beta H_0,[\beta H_0,Q]](s)=(\beta\epsilon)^2 Q(s)  
\label{3.13}
\end{align}
one establishes that $\left\langle \R(|s|)\R\right\rangle_\text{rad}$
obeys the differential equation
\begin{align}
\left[\frac{\partial^2}{\partial s^2}-(\beta\epsilon)^2\right]\left\langle \R(|s|)\R\right\rangle_\text{rad}=\delta(s)
\end{align}
Its solution with periodic boundary condition $\left\langle \R(s=0)\R\right\rangle_\text{rad}=\left\langle \R(s=1)\R\right\rangle_\text{rad}$ is
\begin{equation}
\left\langle \R(|s|)\R\right\rangle_\text{rad}=\frac{\e^{-\beta\epsilon(1-|s|)}+\e^{-\beta\epsilon |s|}}{2\beta\epsilon (1-\e^{-\beta\epsilon})}
\label{3.15}
\end{equation}

\subsection{Path integral representation for the electronic variable}

We observe that the configurational matrix element of the electronic operator \eqref{3.10}
\be
\label{echange}
(\br|\langle e^{-\beta H}\rangle_\text{rad}|\br)
=\left\langle (\br|{\cal T}\exp\left[-\beta\int_{0}^{1} ds H_\text{el}(\R(s))\right]|\br)\right\rangle_\text{rad}
\ee
is the field average of the imaginary time propagator associated with the Hamiltonian $H_\text{el}(\R(s))$ (\ref{defHelRs}) of an electron in presence of a classical time-dependent 
magnetic field, with potential vector $\bA(\br,s)=\sqrt{2 \beta\epsilon}\R(s)\;\f(\br)$.
We recall that
 the Feynman-Kac-It\^o representation of the configurational diagonal matrix element of this propagator
 \cite{Feynman-Hibbs1965}, \cite{Roepstorff1994}, \cite{Simon1979} reads
\bea
&&( \br |
   {\cal T}\exp\left[-\beta\int_{0}^{1} ds H_\text{el}(\R(s))\right]|\br)\!=\!
   \frac{1}{\(2\pi\lambda^2\)^{3/2}} \!\!\!\int\!\!\! D[\bxi]\nonumber\\
    &&\times
 \exp\left(-\beta\int_{0}^{1}\!\!\! d s\ V\big(\br(s)\big)\right)    
 \exp\left(\moni\tfrac{e\lambda\sqrt{2 \beta\epsilon}}{\hbar c}\int_{0}^{1} \!\!\! d\bxi(s) \cdot
 \f\big(\br(s)\big)\;  \R(s) \right)
\label{FKI}
\eea
Here $\br(s)=\br+\lambda\bxi (s)$ is a closed electronic path starting and ending in $\br$. It is written in terms of
a closed dimensionless Brownian path  at the origin $\bxi(s)
,\;0\leq s \leq 1,\; \bxi (0)=\bxi (1)=\mathbf{0}$ (a Brownian bridge)
and $\lambda=\hbar\sqrt{\frac{\beta}{m}}$ is the thermal de Broglie length of the electron. The measure
$D[\bxi]$ is the corresponding conditional
Wiener measure normalized to $1$. This measure is Gaussian, formally written as
\be
D[\bxi]=\exp\Big(-\frac{1}{2}\int_0^1 d s \left|\frac{d\bxi (s)}{ds}\right|^2\Big)
   d[\bxi (\cdot)]\;\;.
\la{3.17}    
\ee     
It has zero mean and covariance
\begin{align}
    \int\!\! D[\bxi]\,\xi^{\mu}(s)\xi^{\nu}(s')=\delta_{\mu\nu}(\min(s,s')-s s')\;,
    \label{3.18}
\end{align}
where $\xi^{\mu}(s)$ are the Cartesian coordinates of $\bxi(s)$. In this
representation a quantum point charge looks like a classical charged closed filament $\br(\cdot)=(\br,\;\bxi)$
located at $\br$ and with a random shape $\bxi(s),\;0 \leq s\leq 1$,
the latter having a spatial extension given by the thermal de Broglie length 
(the typical size of quantum position fluctuations).  
The magnetic phase in \eqref{FKI} is a stochastic line integral: it is the flux of the magnetic field
across the closed filament.

From the Feynman-Kac-It\^o formula \eqref{FKI}, the average over the radiation degrees of freedom to be 
performed in \eqref{echange} occurs in the form of the Fourier transform of the Gaussian measure whose covariance  is $\left\langle \R(s)\R(s')\right\rangle_\text{rad}$. Hence from the basic formula 
\begin{equation}
\left\langle\exp\left(\moni \int_{0}^1ds \,a(s)\R(s)\right)\right\rangle_\text{rad}
=\exp\left(-\frac{1}{2}\int_{0}^1ds\int_{0}^1ds'\,a(s)\left\langle \R(s)\R(s')\right\rangle_\text{rad}a(s')\right)
\label{3.21}
\end{equation}
for the Fourier transform of a Gaussian,
the matrix element \eqref{echange} eventually  reads
\bea
\label{ToutF}
&&(\br|\langle e^{-\beta H}\rangle_\text{rad}|\br)
=  \nonumber\\&& \frac{1}{\(2\pi\lambda^2\)^{3/2}}
 \;\int\! D[\bxi] \exp\left(-\beta\int_{0}^{1}\!\!\! d s\ V\big(\br(s) \big)\right)    
 \exp\left(-\frac{\beta e^2}{2}\cW_\text{rad}\[\br,\bxi\])\right) 
\eea
with
\be
\label{defWgen}
\cW_\text{rad}\[\br,\bxi\]=\frac{2\lambda^2 \epsilon}{ \hbar^2 c^2}
\int_0^1 d\xi^\mu(s)\int_0^1 d\xi^\nu(s')
\left\langle \R(s)\R(s')\right\rangle_\text{rad} f^\mu\big(\br(s) \big)
f^\nu\big(\br(s') \big)
\ee
This formula can be readily extended to the case where the field has a multimode expansion (\ref{3.1b}) with mode variables $Q_{\gamma},P_{\gamma}$ and mode energies $\epsilon_{\gamma}$. Reinstalling the mode indices and noting that the modes are independent and
identically distributed, i.e.
$\langle \R_{\gamma}(s)\R_{\gamma'}(s')\rangle_\text{rad}=\delta_{\gamma\gamma'}\langle \R_{\gamma}(s)\R_{\gamma}(s')\rangle_\text{rad}$, Eq.\eqref{defWgen} becomes
\be
\label{multimode}
\cW_\text{rad}\[\br,\bxi\]=\frac{2\lambda^2}{ \hbar^2c^2}
\sum_{\gamma}\int_0^1 d\xi^\mu(s)\int_0^1 d\xi^\nu(s')\epsilon_{\gamma}
\left\langle \R_{\gamma}(s)\R_{\gamma}(s')\right\rangle_\text{rad} f^\mu_{\gamma}\big(\br(s) \big)
f^\nu_{\gamma}\big(\br(s') \big)
\ee
Eq. \eqref{ToutF} is the desired generalization of the Feynman-Kac formula when the electron is not only submitted to a static potential $V$ but also to the field-induced potential $\cW_\text{rad}$ mediated
by the presence of the photon field.

\section{Atom near a metallic wall}

\subsection{Description of the system}

The system consists of a Hydrogen atom  located in the vicinity of a metallic wall, the surface of which is located at $x=0$, and interacting with a quantum electromagnetic field. The
atom is made of an infinitely heavy proton at $\bR=(X,0,0),\;X>0,$ and of a quantum electron  of mass $m$ and charge $e$ at $\br=\bR+\ba$, $\ba$ being the relative position of the electron with respect to that of the proton.
The wall is treated macroscopically : on the wall surface at $x=0$, the electronic wave function is assumed to satisfy Dirichlet boundary conditions while the electromagnetic field satisfies the metallic boundary conditions. The electrostatic interaction with the wall is determined by the method of image charges. 

Thus the electrostatic potential part of the Hamiltonian (\ref
{3.1})
\begin{align}
V({\br})= -\frac{e^2}{|\br-\bR|}-e^2V_\text{im}(\br)
\label{1.1}
\end{align}
is the sum of the Coulomb interaction between the electron and the proton in the bulk and of their interactions with image charges located at $\bR^*=(-X,0,0)$ for the proton and $\br^*=\bR^*+\ba^*, \ba^*=(-a_{x},a_{y},a_{z})$ for the electron:
\be 
V_\text{im}(\br)=\frac{1}{2}\(\frac{1}{|\bR-\bR^*|}+\frac{1}{|\br-\br^*|}-\frac{1}{|\bR-\br^*|}-\frac{1}{|\br-\bR^*|}\)
\label{defVim}
\ee
With the definition of the Fourier transform
\be
f(X)= \int \frac{d\bk}{(2\pi)^3} e^{-\moni k_x 2X} \widetilde{f}(\bk)
\label{defFourier}
\ee
we can write
\begin{align}
\label{defVimFourier}
&V_\text{im}(X) = \int \frac{d\bk}{(2\pi)^3} e^{-\moni k_x 2X}\widetilde{V}_\text{im}(\bk)\nonumber\\
&\widetilde{V}_\text{im}(\bk)=\frac{4\pi}{k^2}\frac{1}{2}\[1+\e^{-\moni\bk\cdot(\ba-\ba^*)}
-\e^{\moni\bk\cdot a^*}-\e^{-\moni\bk\cdot a}\]
\end{align}
The large distance asymptotics of $V_\text{im}(X) $ is determined by the behaviour of its Fourier transform as $\bk\to 0$,
which reads from (\ref{defVimFourier})
\begin{align}
\widetilde{V}_\text{im}(\bk)\sim\frac{2\pi}{k^2}(\bk\cdot\ba)(\bk\cdot\ba^*) 
\label{VimFourierasymptotic}
\end{align}
This implies that the atom and its image charge have    the dipolar interaction
\begin{align}
e^2V_\text{im}(X)\sim\frac{e^2}{2}\left[\frac{\ba\cdot\ba^*-3(\hat{\bR}\cdot\ba)(\hat{\bR}\cdot\ba^*)}{R^3}\right]_{\bR=(2X,0,0)}
=\frac{e^2}{(2X)^3}\left[a_{x}^2+\tfrac{1}{2}(a_{y}^2+a_{z}^2)\right]
\label{Vimspaceasymptotics}
\end{align}
as $X\to \infty$.
\bigskip

The electromagnetic field  is enclosed in a metallic box $L\times L\times L$ whose face at $x=0$ coincides with the wall position. This gives rise to the eigenmode expansion of the vector potential
of the form (\ref{3.1b}) with mode indices $\gamma=(\bk,\eta)$
\begin{equation}
\label{decompA}
A^\mu(\br)=\sqrt{\frac{16\pi\hbar c}{ L^3}}\sum_{\bk,\eta}\frac{g(\lcut k)}{\sqrt{k}}
\e_\eta^\mu(\bk)(a_{\bk\eta}+a^{\dag}_{\bk\eta})h_\mu(\bk,\br)
,\quad \mu=x,y,z
\end{equation} 
where $\bk$ are wave numbers and  $\boe_\eta^\mu(\bk),\eta=1,2$, are two unit polarization vectors orthogonal to $\bk$. The functions $h_\mu(\br)$ are the cavity
modes corresponding to a transverse electric field whose longitudinal part vanishes on the metallic faces of the box at $x=0$, $x=L$, $y=0$, $y=L$, $z=0$ and $z=L$:
\begin{eqnarray}
&&\qquad h_x(\bk,\br)=\cos(k_xx)\sin(k_yy)\sin(k_zz)\nonumber\\
&&\qquad h_y(\bk,\br)=\sin(k_xx)\cos(k_yy)\sin(k_zz)\nonumber\\
&&\qquad h_z(\bk,\br)=\sin(k_xx)\cos(k_yy)\cos(k_zz),\nonumber\\
&& k_x=\frac{\pi n_x}{L}, k_y=\frac{\pi n_y}{L},\quad k_z=\frac{\pi n_z}{L},\quad n_x,n_y, n_z=0,1,2,\ldots
\la{1.3}
\end{eqnarray}
In \eqref{decompA} $g(\lcut k)$ is a smooth spherically symmetric ultraviolet cut-off, $g(0)=1$, and $g(\lcut k)=0$ when $k>\lcut^{-1}=mc/\hbar$. The prefactor in \eqref{decompA} includes the normalization
of the eigenmodes \eqref{1.3} and assures the correct form of the free radiation Hamiltonian \eqref{3.1c} with eigenmode energies 
\begin{equation}
\epsilon_{\gamma}\equiv \epsilon_k= \hbar \omega_k,\;\quad\omega_k=ck
\label{1.4}
\end{equation}
Note that in this model we have not included the spin of the electron and its Pauli coupling with the electromagnetic field.

\subsection{Effective atom-wall interaction}

Assuming that the electron and the photons are in thermal equilibrium at temperature $T$, we consider  the immersion  free energy of the atom  in the photon field  and in presence of the wall 
\begin{equation}
\label{defImmersion}
F(X, \beta)=-k_B T \lim_{L\to\infty}
\(\ln\frac{\Tr^{\scriptscriptstyle\text{L}}\e^{-\beta H}}{\Tr^{\scriptscriptstyle\text{L}}_\text{rad}\e^{-\beta H_\text{rad}}}\)
\end{equation}
where the trace $\Tr^{\scriptscriptstyle\text{L}}=\Tr^{\scriptscriptstyle\text{L}}_\text{el}\Tr^{\scriptscriptstyle\text{L}}_\text{rad}$  runs over the space of electronic wave functions together with the states of the photonic Fock space which satisfy the boundary conditions on the box surfaces. The limit $L\to\infty$ means that the field region has been extended to the whole half-space $x\geq 0$. 
By virtue of \eqref{Vimspaceasymptotics} $V_\text{im}$ goes to zero when $X$ goes to infinity ; henceforth, after the  limit $L\to \infty$ has been taken, the bulk quantities can be obtained from the corresponding ones in the wall vicinity by taking the limit
 of an infinite $X$.
 Then the effective atom-wall interaction 
\begin{align}
\Phi(X,\beta)=F(X, \beta)-\lim_{X\to\infty}F(X, \beta)
\label{defPhi}
\end{align}
is defined as the difference
between the immersion free energy $F(X, \beta)$ when the atom is located at distance $X$ from the wall and its corresponding bulk value obtained as $X\to\infty$. After inserting the definition (\ref{defAverageRad}) into  \eqref{defImmersion} and performing the electronic trace in the configuration representation, $\Phi(X,\beta)$ may be expressed as
\bea
\label{defPhiBis}
&&\Phi(X, \beta)=-k_B T
\\\nonumber
&&\quad\times  
\ln \(\int d\br (\br\vert \langle e^{-\beta H(X)}\rangle_\text{rad}\vert \br)/\lim_{X\to +\infty}\int d\br (\br\vert \langle e^{-\beta H(X)}\rangle_\text{rad}\vert \br) \)
\eea
It is understood here that the field extends in the half-space $x\geq 0$ and the notation $H(X)$ recalls that the Hamiltonian depends on the atom-wall distance $X$.
Our main interest will be the analysis of $\Phi(X,\beta)$ as a function of the distance $X$ for various temperature regimes.

At this point we repeat the warning already given in the introduction about diverging atomic traces. The spatial integrals
on electronic configurations in (\ref{defPhiBis}) must be restricted
to some finite spatial region representing the effective available space for the atom in a low density phase.   

\subsection{Functional expression of the effective atom-wall interaction}

In view of (\ref{defPhiBis}) we have to specialize the generalized Feynman-Kac formula \eqref{ToutF} to the atom-wall model. The electrostatic potential takes the standard path integral form
\begin{align}
V(\br,\bxi)= \int_{0}^1dsV(\br(s))= -e^2
\int_{0}^1ds\frac{1}{|\br(s)-\bR|}-e^2\int_{0}^1dsV_\text{im}(\br(s))
\label{3.3}
\end{align}
whereas the effective magnetic potential $\cW_\text{rad}^{\scriptscriptstyle\text{L}}(\br,\bxi)$ in the finite box $L^3$ becomes
\begin{eqnarray}
\label{defWmL}
\cW_\text{rad}^{\scriptscriptstyle\text{L}}\[\br,\bxi\]&=&\frac{\lambda^2}{\lambda_{\text{ph}}^2 }\frac{8}{L^3}\sum_{\bk}\delta_{\mu\nu}^{\perp}(\bk)
\frac{4\pi g^2(\lcut k)}{k^2}\int_{0}^1
d\xi^\mu(s)\int_{0}^1d\xi^\nu(s')\nonumber\\ &\times&
\Q(\lambda_{\text{ph}}k,s-s')h_{\mu}(\bk,\br(s))h_{\nu}(\bk,\br(s'))
\end{eqnarray}
This follows from a comparison between (\ref{3.1b}) and \eqref{decompA} showing that the mode functions in \eqref{multimode} (including the proper multiplicative factors) have to be identified with\
\begin{align}
f^{\mu}_{\bk,\eta}(\br)= \sqrt{\frac{16\pi\hbar c}{L^3}} \frac{g(\lcut k)}{\sqrt{k}}e^ \mu_{\eta}(\bk)h^{\mu}(\br)
\label{3.4}
\end{align}
In \eqref{defWmL}
\begin{equation}
\delta_{\mu\nu}^{\perp}(\bk)=\sum_{\eta}e_{\eta}^{\mu}(\bk) e_{\eta}^{\nu}(\bk)
=\delta_{\mu\nu}-\frac{k_\mu k_\nu}{k^2}
\label{transversedelta}
\end{equation}
is the transverse Kronecker symbol resulting from the summation on polarization indices and it is convenient to make the definition
\be
\Q(\lambda_{\text{ph}}k,s)=(\beta \epsilon_k)^2\langle \R_{\bk}(|s|)\R_{\bk}\rangle_{\text{rad}}
\ee
By virtue of \eqref{3.15}
\begin{align}
&\Q(\lambda_{\text{ph}}k,s)=
\frac{\lambda_{\text{ph}}k}{2}\frac{\e^{-\lambda_{\text{ph}}k(
1-|s|)}+\e^{-\lambda_{\text{ph}}k|s|}}{1-\e^{-\lambda_{\text{ph}}k}}\nonumber\\
&\lambda_{\text{ph}}k=\beta \epsilon_k=\beta\hbar ck
\label{defQ}
\end{align}
The function ${\cal Q}(u,{s})$ is the thermal photon propagator.
As seen from (\ref{defQ}) it is the Green function
of the quantum harmonic oscillator with periodic boundary conditions ${\cal Q}(u,0)={\cal Q}(u,1)$.
It satisfies the normalizations
\begin{align}
\int_{0}^1ds {\cal Q}(u,{s})=1,\quad\lim_{u\to 0} {\cal Q}(u,{s})=1
\label{3.26}
\end{align}
One has ${\cal Q}(u,{s})\geq 0$, ${\cal Q}(u,1-{s})={\cal Q}(u,{s}),\;0\leq s\leq 1$ and ${\cal Q}(u,{s})$
can be extended to a periodic function of $s$
of period $1$ for all $s$. 
\bigskip

It remains to extend the box $L^3$ to the whole half-space $x\geq 0$. As a result, $\lim_{L\to\infty}\cW_\text{rad}^{\scriptscriptstyle\text{L}}(\br,\bxi)=\cW_\text{rad}(x,\bxi)$ will become independent of the location of the electronic filament $\bxi$ in the $y,z$ directions. So we can equivalently perform a spatial average
$\frac{1}{L}\int^{L}_{0}dy\frac{1}{L}\int^{L}_{0}dz\l=\overline{\rule{0pt}{6pt}\l}$,
replacing $\overline{\sin^2(k_{y}y)}=1/2,\;\overline{\sin^2(k_{z}z)}=1/2,\;\overline{\cos^2(k_{y}y)}=1/2,\;\overline{\cos^2(k_{z}z)}=1/2$ when developing the products \\$h_{\mu}(\bk,\br(s))h_{\nu}(\bk,\br(s'))$ in \eqref{defWmL}. We find for instance
\begin{align}
&\overline{h_{x}(\bk,\br(s))h_{x}(\bk,\br(s'))}=
\nonumber\\
&\frac{1}{4}
\cos k_{x}x(s)\cos k_{x}x(s')\cos k_{y}[y(s)-y(s')]
\cos k_{z}[z(s)-z(s')]
\label{3.27}
\end{align}
Noting that $$\cos k_{x}x(s)\cos k_{x}x(s')=\tfrac{1}{2}\cos k_{x}[2x+\lambda \xi_{x}(s)+\lambda\xi_{x}(s')]+
\tfrac{1}{2}\cos k_{x}\lambda[\xi_x(s)-\xi_x(s')]$$
we see that the second term in the r.h.s is $x$-independent.When $L$ goes to infinity, the discrete sum on modes $\frac{1}{L^3}\sum_{\bk}$ where $k_x=\pi n_x/L$, $k_y=\pi n_y/L$ and $k_z=\pi n_z/L$ with $ n_x,n_y, n_z=0,1,2,\ldots$ is replaced by integrals
$\int_{0}^{+\infty} \frac{dk_x}{\pi}\int_{0}^{+\infty} \frac{dk_y}{\pi}\int_{0}^{+\infty} \frac{dk_z}{\pi}$. Noting that
$
\deltaT_{xx}(\bkhat)=\(1-\frac{k_x^2}{\bk^2}\)
$ and the remainder of the integrand are even  functions of $k_{x}$, $,k_{y}$ and $k_{z}$
the integration can be extended over the whole of $\bk$ space, e.g.
$$
\int_{0}^{\infty}\frac{dk_{x}}{\pi}\int_{0}^{\infty}\frac{dk_{y}}{\pi}\int_{0}^{\infty}\frac{dk_{z}}{\pi}
\cos(k_x A_x)\ldots=\int\frac{d\bk}{(2\pi)^3}\e^{-\moni k_{x}A_x}\ldots
$$
 and eventually
\bea
&& \lim_{L\to\infty}\frac{8}{ L^3}\sum_{\bk}
\frac{g(\lcut k)}{k^2}\deltaT_{xx}(\bkhat)\overline{h_x\(\bk,\br(s)\) h_x\(\bk,\br(s'))\)}
\\
\nonumber
&=&\int \frac{d\bk}{(2\pi)^3}
\frac{g(\lcut k)}{k^2}\deltaT_{xx}(\bkhat)
\[e^{-\moni k_x[2x +\lambda\xi_x(s)+\lambda\xi_x(s')]}+e^{-\moni k_x\lambda[\xi_x(s)-\xi_x(t)]}\]
\\
\nonumber
&&\qquad\qquad\qquad \times e^{-\moni k_y\lambda[\xi_y(s)-\xi_y(t)]-\moni k_z\lambda[\xi_z(s)-\xi_z(t)]}
\eea
The second term in the r.h.s is $x$-independent and thus contributes to the bulk value $\cW_\text{rad}^\text{b}(\bxi)=\lim_{x\to\infty}\cW_\text{rad}(x,\bxi)$ of the field-induced potential.
Dealing in a similar way with the other components, we  eventually find
\begin{align}
\cW_\text{rad}\[x,\bxi\]=\cW_\text{rad}^\text{b}\[\bxi\]+\cW_\text{rad}^\text{w}\[x,\bxi\]
\label{3.28}
\end{align}
where 
\begin{align}
\cW_\text{rad}^\text{b}\[\bxi\]
&=\frac{\lambda^2}{\lambda^2_{\text{ph}}}
\int\frac{d\bk}{(2\pi)^3}\delta^{\perp}_{\mu\nu}(\bk)
\frac{4\pi g(\lcut k)}{k^2}\int_{0}^1
d\xi^\mu(s)\int_{0}^1d\xi^\nu(s')\nonumber\\
&\times\e^{-\moni\lambda\bk\cdot[\bxi(s)-\bxi(s')]}\Q(\lambda_{\text{ph}}k,s-s')
\label{3.29}
\end{align}
represents the bulk self-energy of the electron in the photon field.
The expression of $\cW_\text{rad}^\text{w}(x,\bxi)$ written in Fourier representation is
\begin{align}
\cW_\text{rad}^\text{w}\[x,\bxi\]=\int \frac{d \bk}{(2\pi)^3}
\e^{-2\moni k_{x}X}\[\frac{4\pi g(\lcut k)}{k^2}\sum_{\mu\nu}
\delta^{\perp}_{\mu\nu}(\bk)\zeta_{\mu\nu}I^{\mu\nu}\[\balpha,\bk\]\]
\label{3.30}
\end{align}
where $I^{\mu\nu}\[\balpha,\bk\]$ is a functional of the position $\balpha(s)=\ba+\lambda\bxi(s)$ of the electron relative to the proton 
\begin{align}
&&I^{\mu\nu}\[\balpha,\bk\]=\frac{1}{\lambda_{\text{ph}}^2}\int_{0}^1
d \alpha^\mu(s)\int_{0}^1d \alpha^\nu(s')\Q(\lambda_{\text{ph}}k,s-s')
\nonumber\\
&&\times\e^{-\moni k_{x}[\alpha_{x}(s)+\alpha_{x}(s')]}\e^{-\moni k_{y}[\alpha_{y}(s)-\alpha_{y}(s')]}\e^{-\moni k_{z}[\alpha_{z}(s)-\alpha_{z}(s')]}
\label{defIm}
\end{align}
and $\zeta_{xx}=1,\;\zeta_{\mu\nu}=-1$ otherwise.
In \eqref{defIm}  we have used the properties $\lambda d\xi_\mu(s)=d \alpha_\mu(s)$ and
$\lambda[\bxi(s)-\bxi(s')]=\balpha(s)-\balpha(s')$.

Collecting these results in (\ref{ToutF}), the explicit form of the generalized Feynman-Kac formula for the atom-wall system becomes
\begin{align}
(\br|\langle e^{-\beta H(X)}\rangle_\text{rad}|\br)&=
  \(\frac{1}{2\pi\lambda^2}\)^{3/2} \!\!\!\int\!\!\! D(\bxi) 
  \exp\[\beta e^2\int_{0}^1ds\tfrac{1}{|\balpha(s)|}
   -\tfrac{\beta e^2}{2}\cW_\text{rad}^\text{b}\[\bxi\]\]\nonumber\\
  &\times\[ \theta(X +\alpha_x(.))\exp\left(\beta e^2
V_\text{im}\[X, \balpha\]
-\tfrac{\beta e^2}{2}\cW_\text{rad}^\text{w}\[X,\balpha\]\right)\]
\label{3.32}
\end{align}
where
\begin{equation}
V_\text{im}\[X, \balpha\]=\int_0^1 ds V_\text{im}\(X, \balpha(s)\)
\label{3.32a}
\end{equation}
is  the standard Feynman-Kac representation corresponding to \eqref{defVim}.
The function $ \theta(u)=1$ if $u\geq 0, \theta(u)=0$ if $u<0$ implements the Dirichlet boundary condition for the electron wave function at the wall.  The second square bracket incorporates the effects of the wall and tends to 1 as $X\to\infty$. Hence
\begin{align}
&\lim_{X\to\infty}(\br|\langle e^{-\beta H(X)}\rangle_\text{rad}|\br)\nonumber\\&=
 \(\frac{1}{2\pi\lambda^2}\)^{3/2} \!\!\!\int\!\!\! D(\bxi) 
  \exp\[\beta e^2\int_{0}^1ds\tfrac{1}{|\balpha(s)|}
   -\tfrac{\beta e^2}{2}\cW_\text{rad}^\text{b}\[\bxi\]\]   =(\br|\langle e^{-\beta H_{\text{b}}}\rangle_\text{rad}|\br)
\label{3.33}
\end{align}
is the generalized Feynman-Kac formula for an atom immersed in an infinitely extended electromagnetic field in homogeneous space, with corresponding Hamiltonian $H_{\text{b}}$. 

In fact the bulk self-energy $\cW_\text{rad}^\text{b}(\bxi)$ of the electron in the homogeneous photon field is identical to that derived in \cite{BuenzliEtAl2007} , formula (66), with the use of periodic boundary conditions.
In other words, the generalized  Feynman-Kac formula in the bulk
Eq.\eqref{3.33} could of course have been established directly starting from the simpler Hamiltonian
$H_{\text{b}}$ obtained by removing image charges and metallic boundary conditions at the wall right away. 
(In the  method used in  Ref.\cite{BuenzliEtAl2007} for a quantum field in the absence of any metallic boundary, the photonic trace is expressed in the basis of  the coherent states associated with the photonic modes, instead of the basis of photonic modes themselves, and then every corresponding matrix element of the Gibbs factor is  replaced by a bosonic path integral different from that of the free oscillator process. However the result of the trace  is the same, as it should.) The spectral properties of this 
Hamiltonian (called The Standard Model of Nonrelativist QED) have been extensively studied (see the recent paper  \cite{Sigal2009} and references therein). $H_{\text{b}}$ has an unique and non degenerate ground state, but atomic excited states are turned into resonances because of the phenomenon of spontaneous emission implying that the rest of the spectrum is continuous.

 Let us introduce the normalized atomic measure in the bulk associated with  $H_{\text{b}}$
\begin{align}
 \int D^\text{b}_\text{at}\[\balpha\]\cdots=
\frac{
\int d \ba \int D[\bxi] 
\exp\[\beta e^2\int_{0}^1ds \tfrac{1}{|\balpha(s)|}
 -\tfrac{\beta e^2}{2}\cW_\text{rad}^\text{b}\[\bxi\]\]\;\;\cdots}
{\int d \ba \int D[\bxi] \exp\[\beta e^2\int_{0}^1ds\tfrac{1}{|\balpha(s)|}
  -\tfrac{\beta e^2}{2}\cW_\text{rad}^\text{b}\[\bxi\]\]}
\label{atomicmeasure}
\end{align}
As discussed in the introduction, the  $\int D^\text{b}_\text{at}\[\balpha\]$ integration needs a regularization by limiting the available space for the electron position in the $d\ba$ integral to a
finite sphere. This regularization will be understood in $D^\text{b}_\text{at}\[\balpha\]$ integrals without additional notation. 
Then, from (\ref{defPhiBis}), (\ref{3.32}) and (\ref{atomicmeasure}) the atom-wall effective potential receives its final form (for a fixed temperature $T>0$)
\begin{align}
\label{Phifinal}
\Phi(X)=
-\kB T \ln  \int D^\text{b}_\text{at}\[\balpha\]\theta(X +\alpha_x(.))\exp\left(\beta e^2 V_\text{im}\[X, \balpha\]-\tfrac{\beta e^2}{2}\cW_\text{rad}^\text{w}\[X,\balpha\]\right) 
\end{align}

Since both $V_\text{im}\[X, \balpha\]$ 
and $\cW_\text{rad}^\text{w}\[X,\balpha\]$ tend to zero as $X\to\infty$, the large distance behaviour of the atom-wall potential  is obtained by expanding (\ref{Phifinal}) to first order in these potentials giving three contributions
\begin{align}
\Phi(X)
\underset{X\to \infty}\sim\Phi_\text{im}(X)+\Phi_\text{rad}(X)+\Phi_{\text {geomc}}(X)
\label{defPhias}
\end{align} 
The two first ones
\begin{align}
&
\Phi_\text{im}(X)= -e^2 \int D^\text{b}_\text{at}[\balpha]V_\text{im}[X,\balpha]\label{defPhiasa}\\
& \Phi_\text{rad}(X)= \frac{e^2}{2} \int D^\text{b}_\text{at}[\balpha]\cW_\text{rad}^\text{w}\[X,\balpha\]
\label{defPhiasb}
\end{align}
arise from the image-charges and the photon field. The last one
\begin{equation}
\label{defPhiasc}
\Phi_{\text{geomc}}(X)=-\kB T \ln  \int D^\text{b}_\text{at}\[\balpha\]\theta(X +\alpha_x(.))
\end{equation}
comes from the pure geometrical constraint imposed
by the wall on the atomic weight. In view of the fact that $V_\text{im}[X,\balpha]$ and $\cW_\text{rad}^\text{w}\[X,\balpha\]$ vanish at large distance, this constraint can be disregarded
in (\ref{defPhiasa}) and (\ref{defPhiasb}) when looking for the dominant term as $X\to \infty$. In the low temperature regime considered in section 4.3, where only ground state contributions of the Hydrogen atom will be kept, $\Phi_{\text{geomc}}(X)$ will be exponentially small for $X\gg \aB$ since the ground state is exponentially localized in the proton vicinity.

\section{Single atom in a thermalized quantum electromagnetic field}

\subsection{The net interaction}

We examine the large distance behaviour of 
the Coulomb potential due to image charges (\ref{defPhiasa}) 
\begin{align}
\Phi_\text{im}(X)=\int \frac{d\bk}{(2\pi)^3} e^{-\moni k_x 2X} \widetilde{\Phi}_\text{im}(\bk)
\label{PhicFouriertransforma}
\end{align}
and of the field-induced potential (\ref{defPhiasb})
\begin{align}
 \Phi_\text{rad}(X)=\int \frac{d\bk}{(2\pi)^3} e^{-\moni k_x 2X} \widetilde{\Phi}_\text{rad}(\bk)
\label{PhimFouriertransform}
\end{align}
written in Fourier representation according to (\ref{defFourier}).
As far as the Coulomb part is concerned, going to the path representation of (\ref{VimFourierasymptotic}) and averaging with the atomic weight $D_\text{at}^\text{b}$, it follows immediately that
\bea
\widetilde{\Phi}_\text{im}(\bk)
&=&-\frac{2\pi e^2}{k^2} \int D_\text{at}^\text{b}[\balpha]\int_0^1 ds\,(\bk\cdot\balpha(s)(\bk\cdot\balpha^*(s))+\cO\(k^2\)\nonumber\\
&=&
\frac{2\pi e^2}{k^2}\sum_{\mu} \zeta_{\mu\mu} k_\mu^2 \int D_\text{at}^\text{b}[\balpha]\int_0^1 ds\,\[\alpha_x(s)\]^2+\cO\(k^2\)
\label{PhicFouriertransformb}
\eea
In the second line, we have used the rotational invariance of the bulk atomic measure to simplify the expression.
Noting that $\sum_{\mu} \zeta_{\mu\mu} k_\mu^2=2k_x^2-k^2$
we conclude that
the leading tail of the Coulomb interaction \eqref{PhicFouriertransforma} arises from the singular part of $\widetilde{\Phi}_\text{im}(\bk)$ which is
\be
\label{AsPhicpath}
\widetilde{\Phi}_\text{im}(\bk)\sim 4\pi e^2\frac{k_x^2 }{k^2}\int D_\text{at}^\text{b}[\balpha]\int_0^1 ds\,\[\alpha_x(s)\]^2. 
\ee

The small $\bk$ behaviour of $\widetilde{\Phi_\text{rad}}(\bk)
$ requires a more elaborate study. According to (\ref{defPhiasb}) and (\ref{3.30}) one has
\be
\widetilde{\Phi}_\text{rad}(\bk)=\frac{4\pi e^2}{k^2}\frac{1}{2}\sum_{\mu,\nu}\deltaT_{\mu\nu}(\bkhat) \zeta_{\mu\nu} \int D_\text{at}^\text{b}[\balpha] I^{\mu\nu}[\balpha,\bk]
\ee
with $I^{\mu\nu}[\balpha,\bk]$ defined in \eqref{defIm}.
The exponentials in $I^{\mu\nu}[\balpha,\bk]$ can be expanded in powers of $\bk$  
\begin{align}
I^{\mu\nu}[\balpha,\bk]=I^{\mu\nu\;[0]}[\balpha,\bk]+I^{\mu\nu\;[2]}[\balpha,\bk]+\cdots
\label{Iexpanded}
\end{align}
where the indices $0,2,\ldots$ refer to exponentials expanded to $0,2,\ldots$ order (odd orders will not contribute because averages of odd powers of $a^\mu$ vanish by rotational invariance of the atomic weight). Accordingly we have the expansion
\begin{align}
&\widetilde{\Phi}_\text{rad}(\bk)=\widetilde{\Phi}_\text{rad}^{[0]}(\bk) + \widetilde{\Phi}_\text{rad}^{[2]}(\bk)+\cdots
\nonumber\\
&\Phi_\text{rad}(X)=\Phi_\text{rad}^{[0]}(X) + \Phi_\text{rad}^{[2]}(X)+\cdots
\label{Phiexpanded}
\end{align}
This does not allow yet to infer the small $\bk$ behaviour of 
$\widetilde{\Phi}_\text{rad}^{[0]}(\bk)$ and $\widetilde{\Phi}_\text{rad}^{[2]}(\bk)$ in a straightforward manner, since there is still a $\bk$ dependence
in the photon thermal function $\Q\(\lph k, |s-t|\)$ and atomic averages have to be performed to obtain these quantities. The following facts turn out to be true:

\begin{itemize}
\item Both $\Phi_\text{rad}^{[0]}(X)$ and $\Phi_\text{rad}^{[2]}(X)$ contribute at the same leading order as $X\to\infty$. 
\item The term $\Phi_\text{rad}^{[0]}(X)$ corresponds to making the dipolar approximation, i.e. neglecting the phase occuring in the exponentials in (\ref{defIm}). This approximation is valid in an atomic state when the photon wave length is much larger than the Bohr radius, i.e. $k\gg 1/\aB$, or equivalently $X\gg \aB$.
\item
The next correction $\widetilde{\Phi}_\text{rad}^{[2]}(X)$ involves the diamagnetic susceptibility of the atom.
Asymptotic contributions of $\Phi_\text{rad}^{[2]}(X)$ are either $\lambda_{\scriptscriptstyle{C}}/X$, with $\aB\ll X\ll \lat$, or $\afs^2$    smaller than those of $\Phi_\text{rad}^{[0]}(X)$ \cite{CornuMartin}, where $\afs$ is the fine structure constant and $\lambda_{\scriptscriptstyle{C}}$ is the Compton length $\lambda_{\scriptscriptstyle{C}}=\hbar/mc$ with $\lambda_C/\aB=\afs$.
\end{itemize}

In the rest of this paper, we shall work in the dipolar approximation, restricting our attention to the first term $\Phi_\text{rad}^{[0]}(X)$.
Consequently, from now on, our total potential will be restricted to the form
\begin{equation}
\Phi^{[0]}(X)=\Phi_\text{im}(X)+\Phi_\text{rad}^{[0]}(X)
\label{totalpotential}
\end{equation}
We shall present the analysis of the finer corrections arising from $\Phi_\text{rad}^{[2]}(X)$ in another paper \cite{CornuMartin}.
\bigskip

Replacing the exponentials by 1 in (\ref{defIm}), we have
\begin{align}
I^{\mu\nu\;[0]}\[\balpha,\bk\]=\frac{1}{\lambda_{\text{ph}}^2}\int_{0}^1
d \alpha^\mu(s)\int_{0}^1 d \alpha^\nu(s')\Q(\lambda_{\text{ph}}k,s-s')
\label{}
\end{align}
where $\Q(\lambda_{\text{ph}}k,s-s')$ is in fact a function of $\vert s-s'\vert$.
Rotational invariance of the measure $D_\text{at}^\text{b}[\balpha]$ implies that 
$\int D_\text{at}^\text{b}[\balpha]\, I^{\mu\nu}[\balpha,\bk]^{[0]}=\delta_{\mu,\nu}I(k)$ with
\be
\label{defI}
I(k)=\frac{2}{\lph^2}\int D_\text{at}^\text{b}[\balpha]\int_0^1 d \alpha^x(s) \int_0^sd \alpha^x(s')\Q\(\lph k, s-s'\)
\ee
where we have used the symmetry of the integrand under the exchange of  $s,s'$ to reduce the time integration to the sector $s'\leq s$.
From the definitions (\ref{defPhiasb}), (\ref{PhimFouriertransform}) and (\ref{3.30}), and in view of the relation
$\sum_{\mu}\zeta_{\mu\mu} \(1-k_\mu^2/k^2\) =-2k_x^2/k^2$, we get eventually
\be
\label{Phimzero}
\widetilde{\Phi}_\text{rad}^{[0]}(\bk)
=- 4\pi e^2\frac{k_x^2}{k^2}\frac{I(k) }{k^2}
\ee
All the relevant information is now contained in the function $I(k)$.

Then one can get rid of the stochastic integrals by means of a double integration by parts.
The calculation can be formally performed by using the standard rules 
for ordinary differentials. (A more detailed  justification for this calculation will be
found in \cite{CornuMartin}). When using standard integration by parts, one takes into account the fact that for a Brownian brigde $\balpha(1)=\balpha(0)$ and the following properties of the function $f(u)=\Q(\lph k, u)$ arising from the definition \eqref{defQ} :  $f(1-u)=f(u)$ and $f'(1-u)=-f'(u)$ for $0\leq u\leq 1$.
As a result, the expression \eqref{defI} is equal to
\bea
\label{Ikresultpath}
\lph^2 I(k)&=&
- 2 \int D_\text{at}^\text{b}[\balpha]  \Q'(\lph k,0)\int_0^1 ds\,\[\alpha_x(s)\]^2
\\
\nonumber
&&-2 \int D_\text{at}^\text{b}[\balpha]\int_0^1ds\int_0^s dt  \Q''(\lph k,s-t)\alpha_x(s)  \alpha_x(t)
\eea
The expression \eqref{defQ} of $\Q(k,u)$  leads to
\be
\Q'(k,0)=-\frac{1}{2} \lph^2 k^2 \quad\text{and}\quad   \Q''(\lph k,u)=\lph^2 k^2\Q(\lph k,u),\;\;\; u\geq 0 
\ee
so that 
\be
\frac{I(k) }{k^2}=
 \int_0^1 ds\,\int D_\text{at}^\text{b}[\balpha]\[\alpha_x(s)\]^2
-B(k)
\ee
with
\be
\label{defA1}
B(k)\equiv 2\int_0^1ds\int_0^s dt \Q(\lph k,s-t) \int D_\text{at}^\text{b}[\balpha]\alpha_x(s)  \alpha_x(t)
\ee
Finally inserting the expression $I(k)$  into \eqref{Phimzero} gives 
\begin{align}
\widetilde{\Phi}_\text{rad}^{[0]}(\bk)
=- 4\pi e^2 \frac{k_x^2}{k^2} \int_0^1 ds\,\int D_\text{at}^\text{b}[\balpha]\[\alpha_x(s)\]^2+4\pi e^2\frac{k_x^2}{k^2}B(k)
\label{Phimzeroa}
\end{align}
At this point one observes the remarkable fact that the first term in (\ref{Phimzeroa}) is nothing else, up to the sign, than the singular
small-$\bk$ expression (\ref{PhicFouriertransformb}) of the Coulombic part of the interaction. Consequently, when considering now the total potential (\ref{totalpotential}) the dipolar long-ranged part $\sim X^{-3}$ due to $\Phi_\text{im}(X)$ is exactly compensated. Thus, the total net potential at large distance
\be
\Phi^{[0]}(X)\sim 4\pi e^2\int \frac{d\bk}{(2\pi)^3} e^{-\moni k_x 2X} \,
\frac{k_x^2}{k^2}B(k)
\label{netpotential}
\ee
is entirely determined by the behaviour of the function $B(k)$
(\ref{defA1}). Clearly this function shows an interplay between
the photon thermal propagator and the squared moduli of  the atomic moments,
which will be made explicit in the next section.

\subsection{Asymptotic regimes}

Although we have disregarded diamagnetic contributions,
 there are still  effects depending on the fine structure constant in the expression (\ref{netpotential}) of $\Phi^{(0)}(X)$.
Indeed since our calculation so far is not perturbative with respect to the
coupling constant (the charge of the electron), the thermal weight $D_\text{at}^\text{b}[\balpha]$ (\ref{atomicmeasure}) of the atom
still includes its coupling with the electromagnetic field. When this coupling is switched on, it turns the excited states of the bare hydrogen atom into resonances, with displaced energy levels and finite life times due to the phenomenon of spontaneous emission (see comments after (\ref{3.33})).

We assume that these effects do not modify the power of  the $X$-decay but only bring tiny corrections to its amplitude. Therfore we  neglect them in the subsequent analysis of $\Phi^{(0)}(X)$.
This amounts to drop the effective interaction $\cW_\text{rad}^\text{b}\[\bxi\]$ in (\ref{atomicmeasure}), namely replacing $D^\text{b}_\text{at}\[\balpha\]$ by the bare hydrogen atom weight
\begin{align}
 \int D^\text{b}_\text{at}\[\balpha\]\cdots \rightarrow
\frac{\int d \ba \int D[\bxi] 
\exp\[\beta e^2\int_{0}^1ds \tfrac{1}{|\balpha(s)|}\]\;\;\cdots}
{\int d \ba \int D[\bxi] \exp\[\beta e^2\int_{0}^1ds\tfrac{1}{|\balpha(s)|}\]}\equiv 
\langle  \cdots \rangle_\text{at}
\label{barehydrogen}
\end{align}
Coming back to the operator formulation
\begin{align}
\langle  A \rangle_\text{at}=\frac{1}{\Zat}\Tr\(e^{-\beta H_\text{at}}A\), \quad\Zat=\Tr\(e^{-\beta H_\text{at}} \), \quad H_\text{at}=\tfrac{p^2}{2m}-\tfrac{e^2}{|\br-\bR|}
\label{defaverageat}
\end{align}
is the usual thermal average for an atomic observable $A$. 
It is important to repeat here the warning about  our phenomenological treatment of screening (see the introduction and the remark after (\ref{atomicmeasure})).
The above formulae have to be regularized, either by a cut-off in the spatial $d\ba$ integral occuring in  
(\ref{barehydrogen}) or by an energy cut-off limiting the evaluation of the traces (\ref{defaverageat}) to a finite number of hydrogen eigenstates with maximal energy $E_{\rm max}$. 

With this limitation we can proceed to an explicit calculation of the function $B(k)$ (\ref{defA1}), which now reads
\be
\label{Bop}
B(k)= 2\int_0^1ds\int_0^s dt \langle a_x(s)  a_x(t)\rangle_\text{at}\Q(\lph k,s-t)
\ee
Coming back to operator langage by inverse Feynman-Kac transformation, the atomic fluctuation can be evaluated in a basis of eigenfunctions for the Hamiltonian  $H_\text{at}$
\begin{align}
\label{aa}
\langle a_x(s)  a_x(t)\rangle_\text{at}&=
\frac{1}{\Zat}\Tr\(e^{-\beta H_\text{at}}a_{x}(s)a_{x}(t)\), \quad a_{x}(s)=
e^{s\beta H_\text{at}}a_{x}e^{-s\beta H_\text{at}}
\nonumber\\
&=
\frac{1}{\Zat}\sum_i^{i_\text{\mdseries max}}\sum_{j\neq i}
\vert (i \vert a_x\vert j)\vert^2 e^{-[1-(s-t)]\beta E_i} e^{-(s-t)\beta E_j}
\end{align}
where $i_\text{\mdseries max}$ is the uppest state index such that $E_i$ is equal to the phenomological cut-off 
$E_\text{\mdseries max}$ discussed in the introduction. The  rotational invariance of $H_\text{at}$ enforces that $(i\vert a_x\vert i)=0$ so that the term $j=i$ can be omitted
\footnote{The j-summation on intermediate states runs over the whole spectrum of tha Hydrogen atom including its countinuous part}.
According to the expression \eqref{defQ}  of $\Q(\lph k,s-t)$, we have to calculate
\be
\frac{1}{\Zat}\beta\int_0^1ds\int_0^s dt \[e^{-(s-t) \beta \epsilon_k }+e^{-[1-(s-t)]\beta\epsilon_k}\]
e^{-\beta E_i} e^{-(s-t)\beta (E_j-E_i)}
\ee
with $\epsilon_k\equiv \hbar c k$.
The result of the integration is the sum of two terms $I_{ij} +J_{ij}$
\begin{align}
&I_{ij}=\frac{1}{\Zat}\[\frac{e^{-\beta E_i}}{E_j-E_i+\epsilon_k}+\frac{e^{-\beta( E_i+\epsilon_k)}}{E_j-E_i-\epsilon_k}\]\label{I_{ij}}\\
&J_{ij}=\frac{1}{\beta\Zat}\[\frac{e^{-\beta (E_j+\epsilon_k)}-e^{-\beta E_i}}{\(E_j-E_i+\epsilon_k\)^2}
-\frac{e^{-\beta (E_i+\epsilon_k)}-e^{-\beta E_j}}{\(E_j-E_i-\epsilon_k\)^2}\]=-J_{ji}
\label{J_{ij}}
\end{align}
and therefore
\be
B(k)=\frac{\epsilon_k}{(1- e^{-\beta \epsilon_k})}
\frac{1}{\Zat}\sum_i^{i_\text{\mdseries max}}\sum_{j\not=i}\vert (i\vert a_x\vert j )\vert^2 
(I_{ij} +J_{ij})\label{general B}
\ee
We discuss now various distance dependences in the low-temperature regime.

\subsection{Low-temperature regime}

We specialize now our study to the low temperature regime (\ref{lowtemp}).
In this situation we neglect exponentially decaying contributions
of the order $e^{-\beta(E_{1}-E_{0})}$, which amounts to only keep  in (\ref{general B}) the terms having at least one ground state contribution \cite{Martin1997,ACM2007}, i.e
$\Zat=e^{-\beta E_{0}}\(1+{\cal O}(e^{-\beta(E_{1}-E_{0})}\)$ and 
\begin{align}
&B(k)=\frac{\epsilon_k e^{\beta E_{0}}}{(1- e^{-\beta \epsilon_k})}
\[\sum_{j\not=0}\vert (0\vert a_x\vert j )\vert^2 
(I_{0j} +J_{0j})+\sum_i^{i_\text{\mdseries max}}\vert (i\vert a_x\vert 0 )\vert^2 
(I_{i0} +J_{i0})\]\nonumber\\
& \quad\quad+{\cal O}(e^{-\beta(E_{1}-E_{0})})
\label{lowtemperatureB}
\end{align}
As explained in the introduction, once  only the contributions with at least one state equal to the ground state are retained, the double sums becomes convergent and we can remove the cut off $i_{{\rm max}}$ in the second sum of \eqref{lowtemperatureB}. Then
sums involving the $J$ function compensate each other because of the antisymmetry of $J_{ij}$ (see (\ref{J_{ij}}), and also $\vert (0\vert a_x\vert j )\vert=\vert (j\vert a_x\vert 0 )\vert$, and we are eventually left with
\begin{align}
&B^{\scriptscriptstyle LT}(k)=\sum_{j\neq 0}\vert (0\vert a_x\vert j )\vert^2
G(E_j-E_0, \epsilon_k,\beta)\label{Blowtemp}
\end{align}
where $G(E_j-E_0, \epsilon_k,\beta)$ is the following function of energies and temperature
\begin{align}
G(E_j-E_0, \epsilon_k,\beta)=\frac{\epsilon_k}{(1- e^{-\beta \epsilon_k})}
\[\frac{1}{E_j-E_0+\epsilon_k}+\frac{e^{-\beta\epsilon_k}}{E_j-E_0-\epsilon_k}\]
\label{G}
\end{align}
The superscript $LT$ in (\ref{Blowtemp}) denotes the low-temperature regime where \\$\exp\[-(E_1-E_0)/\kB T\]$ terms are disregarded.

Subsequently, in the $X$-behaviour of $\Phi^{[0]}(X)$ in the latter temperature  regime, it is consistent to neglect the oscillating function of  $2 X/\lat$   with a damping factor $\exp\[-(E_1-E_0)/\kB T\]$ which arises in  the inverse Fourier transform \eqref{netpotential} from the singular term $e^{-\beta\epsilon_k}/(E_j-E_0-\epsilon_k +\moni\times 0^+)$ (where $E_j-E_0>0$ and $\epsilon_k>0$.). We recall that $\lat=\hbar c/(E_{0}-E_{1})$ is the  wavelength of the  photon emitted when the atom jumps from the first excited state.

To distinguish now the various possible large distance tails in the low-temperature limit it is useful  to make the scaling $\bk =\bq/X$ in the Fourier integral (\ref{netpotential}) where $\bq$ is a dimensionless Fourier variable, leading to $
B(k)= B(q/X)$ and $\epsilon_k=\epsilon_q/X$.
Hence in (\ref{G}) we have
\be
E_j-E_0\pm\epsilon_k = E_j-E_0\pm\frac{\hbar cq}{X}
=\frac{\hbar c}{X}\left(\frac{X}{\lat}\pm q\right)
\label{scaling}
\ee
and
\be
 \beta\epsilon_k=\frac{\lph}{X}q
\label{scalingbis}
\ee
Since $\lat/\lph=k_{B}T/(E_{1}-E_{0}) \ll 1$ in the considered low-temperature regime (\ref{lowtemp}), one has to discuss separately the cases when $X$ is much smaller (much larger) than $\lat$ or $\lph$.

\subsubsection{Electrostatic dipole interaction $X\ll \lat\ll\lph$}

According to (\ref{scaling}), when $X\ll \lat$ we can neglect $E_j-E_0$ compared to $\epsilon_k$ in (\ref{G}), so that
\be
G(E_j-E_0, \epsilon_k,\beta)
\underset{1\ll \lat k}{=}1+{\cal O}\(\frac{1}{\lat k}\)
\ee
Subsequently, since $\sum_{j}\vert (0\vert a_x\vert j )\vert^2=(0\vert a_x^2\vert 0 )$, \eqref{Blowtemp} simply yields
\be
\label{Bbetainflat}
B^{\scriptscriptstyle LT}(k)\underset{1\ll \lat k}{=}(0 \vert a_x^2 \vert 0)
\ee
By the inverse Fourier transform (\ref{netpotential})
\be
\label{Phicas}
\Phi^{[0]}(X)\underset{\lat \ll X}{\sim}-\frac{1}{X^3} \frac{1}{4} e^2 (0 \vert a_x^2 \vert 0)\equiv\phi_\text{vdW}(X) 
\ee
Thus at  distances $X\ll \lat$  we recover the standard electrostatic dipolar interaction, which could have been inferred from (\ref{Vimspaceasymptotics}) by replacing there the $a_{\mu}^2$  by mean ground state atomic moments $(0\vert a_\mu^2\vert 0)$.

\subsubsection{Retarded interaction $\lat\ll X\ll\lph$}

When $\lat \ll X$  we can neglect $\epsilon_k$  compared to $E_j-E_0$ in (\ref{G}) 
\be
G(E_j-E_0, \epsilon_k,\beta)
\underset{\lat k \ll 1}{=}
\delta_{E_0,E_j} +\[1-\delta_{E_i,E_j} \] \epsilon_k\coth\(\frac{\beta \epsilon_k}{2}\)
\frac{1}{E_j-E_0}
\[1+{\cal O}\(\lat k\)\]
\label{G1}
\ee
Moreover for $X\ll\lph$  the argument $\beta\epsilon_k/2$ of $\coth$ can be considered to be very large (see (\ref{scalingbis}))
\begin{equation}
 \epsilon_k \coth\(\lph  k\)\underset{1\ll k \lph}{=} \hbar c k  \[1+{\cal O}\(e^{-\lph k}\)\]
\label{coth1}
\end{equation}
and inserting (\ref{coth1}) and (\ref{G1})  in  (\ref{Blowtemp})
\begin{equation}
B^{\scriptscriptstyle LT}(k)\underset{\lat k \ll 1}{=}\hbar c k\sum_{j\neq 0}\frac{\vert (0\vert a_x\vert j )\vert^2}{E_j-E_0}\equiv B_{\text{CP}}(k)
\label{retarded1}
\end{equation}
(The ground state is nondegenerate so that the latter sum involves no singularity.)
The corresponding spatial decay follows from the inverse Fourier transform (\ref{netpotential})
\begin{align}
\Phi^{[0]}(X)\underset{\lat\ll X\ll \lph}{\sim} -\frac{\hbar c}{X^4}\frac{3e^2}{4\pi}\sum_{j\neq 0}\frac{\vert (0\vert a_x\vert j )\vert^2}{E_j-E_0}= -\frac{\hbar c}{X^4}\frac{3e^2}{8\pi}\alpha_{E}\equiv\Phi_{\text{CP}}(X)
\label{retarded2}
\end{align}
where 
\begin{align}
\alpha_{E}=2e^2\sum_{j \neq 0}\frac{\vert (0\vert a_x\vert j )\vert^2}{E_j-E_0}
\label{polarizability}
\end{align}
denotes the static polarizability of the hydrogen atom in its ground state (the response of the atom to a classical external electric field in a given direction).
This is precisely the formula originally found by Casimir and Polder \cite{CasimirPolder1948} for retarded interaction in the atom ground state. Our derivation does not involve explicitly the concept of retardation associated with the propagation of Maxwell waves as in \cite{CasimirPolder1948}. The change in the decay regime from $X^{-3}$ to $X^{-4}$ occurs as a consequence of the  behaviour of the photon thermal propagator $\Q(\lph k,t)$ for large wave number $\bk$ producing the extra $k$ factor in (\ref{coth1}) and (\ref{retarded1}). Note that the original ground state Casimir-Polder formula remains valid as long as one neglects thermal effects ${\cal O}(e^{-\beta(E_{1}-E_{0})})$.

\subsubsection{Classical field regime $\lph \ll X $}

When $X\gg \lph$, $\beta\epsilon_k\ll1$ and we expand $\coth$
for small argument
\begin{equation}
 \epsilon_k \coth\(\lph k\)\underset{ \lph k \ll 1}{=}2 \kB T \[1+{\cal O}\(\lph k\)\]
\label{coth2}
\end{equation}
with the result
\begin{equation}
B^{\scriptscriptstyle LT}(k)\underset{ \lph k \ll 1}{=}2k_{B }T\sum_{j\neq 0}\frac{\vert (0\vert a_x\vert j )\vert^2}{E_j-E_0}\equiv B_{\text{class}}(k)
\label{classical1}
\end{equation}
and
\begin{equation}
\Phi^{[0]}(X)\underset{\lph \ll X}{\sim}-\frac{e^2}{2X^3}k_{B}T\sum_{j\neq 0}\frac{\vert (0\vert a_x\vert j )\vert^2}{E_j-E_0}=-\frac{e^2}{4X^3}k_{B}T\;\alpha_{E}\equiv \Phi_\text{class}(X)
\label{classical2}
\end{equation}
The same result can be obtained by expanding the photon propagator for small $k$. Since $\Q(\lph k,t)$ is an even function of $k$ it behaves as $\Q(\lph k,t)=1 + {\cal O}(k^2)$ (see (\ref{defQ})). The term of order $k^{2n}$ in $B(k)$ \eqref{Bop} gives an analytic contribution to the  integrand in (\ref{netpotential})  and rapidly decaying 
terms as $X\to\infty$.
Thus we can set $\Q(\lph k,t)=1$ in (\ref{Bop}) 
\begin{align}
 B_{\text{class}}(k)=2\int_0^1ds\int_0^s dt \langle a_x(s)  a_x(t)\rangle_\text{at}
\label{Bclass}
\end{align}
or equivalently set $\bk=0$ or $\lph=0$ in subsequent formulae, which leads to 
(\ref{classical2}). Setting $\lph=\beta\hbar c=0$ is the same as treating the field classically by turning off the Planck constant in field expressions, hence (\ref{classical2}) describes the atom-wall force when the atom is immersed in a classical electromagnetic field. This potential vanishes at $T=0$. 

We stress again the subtle behaviour of the force at finite (but low) temperature: three
successive ranges of decays occur, as shown in Fig.1,
$$\sim X^{-3}\;\;\text{if}\;\; X\ll \lat , \quad \sim X^{-4}\;\;\text{if}\;\;\lat\ll X\ll\lph,\quad\sim X^{-3}\;\;\text{if}\;\; X\gg \lph$$ 
Note that the asymptotic formulae (\ref{Phicas}), (\ref{retarded2}) and (\ref{classical2}) are exact in the sense that they do not depend on regularization procedures of Coulombic traces. This is because  in (\ref{lowtemperatureB})
one has only retained contributions of the ground state, which is
localized in space.
\begin{figure}
\includegraphics[width= \textwidth]{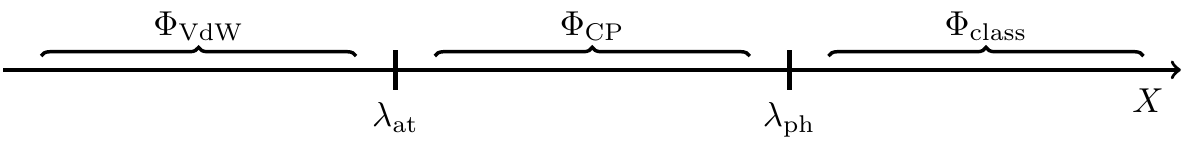}
\caption{Low-temperature atom-wall potential when $\lscreen$ is larger than all  length scales at stake.}
\end{figure}

\section{Screening effects}
\label{ScreeningEffects}

As already said, as soon as the temperature is different from zero, the concept of isolated atom does not make sense. One must rather 
consider a non zero density equilibrium phase of nuclei and electrons: if both the density and the temperature are sufficiently low, the latter can predominantly be found in atomic bound states, but there always remains a fraction of dissociated charges that provide a screening of the bare Coulomb interaction on distances greater than $\lscreen$. This situation is precisely described by the
so called atomic limit that defines the Saha regime (See Ref.\cite{BrydgesMartin1999} (section 7) and  Ref.\cite{ACM2007}). One obtains
another case when the atoms under consideration are also in equilibrium with other species of dissociated charges, like in a strongly ionized gas or an electrolyte. 
The simplest model of the latter case is obtained by embedding the atom in a classical weakly coupled plasma.
Here again Coulomb interactions are screened beyond some typical length $\lscreen$ depending on the plasma density.
In both cases, screening has two main effects on the atomic phase
\begin{itemize}
\item 
Regularization of divergent atomic traces
\item
Reduction of the range of inter-atomic forces
\end{itemize}
Concerning the first point, we just recall here that the regularization consists in substracting
to the Coulomb weight in (\ref{barehydrogen}) a number of terms of its large distance expansion, namely replacing $\exp\[\beta e^2\int_{0}^1ds \tfrac{1}{|\balpha(s)|}\]$ by
\begin{align}
\exp\[\beta e^2\int_{0}^1ds \tfrac{1}{|\balpha(s)|}\]-\sum_{n=0}^N\frac{1}{n\;!}
\[\beta e^2\int_{0}^1ds \tfrac{1}{|\balpha(s)|}\]^n, \quad \balpha(s)=\ba +\lambda\bxi(s) 
\label{truncation}
\end{align}
which is integrable at $\ba=\infty$ whenever $N\geq 3$.
The equivalent truncation in operator language consists in substracting the first terms of the Dyson expansion of the thermal propagator (see (5.11) and (5.12) in \cite{ACM2007} or (117) and (119) in \cite{BalleneggerMartin2003}). This truncation is by no means arbitrary, it follows from the systematic treatment of screening provided by the so called screened cluster expansion presented in \cite{ABCM2003}. Once these truncations have been introduced, it can be rigorously established, following section 5.2 of \cite{ACM2007} or section 6.2 of \cite{BalleneggerMartin2003} and the appendices of these papers, that the remainder in
(\ref{lowtemperatureB}) is indeed ${\cal O}(e^{-\beta(E_{1}-E_{0})})$ (up to a polynomial in $\beta$).

Concerning the second point, the bare Coulomb potential between two charges $e_{1}, e_{2}$ (written in Fourier representation)
\begin{align}
\widetilde{V}_{\scriptscriptstyle C}(\bk,\bxi_{1},\bxi_{2})=\frac{4\pi e_{1}e_{2}}{k^2}\int_{0}^1ds
\e^{\moni\bk\cdot(\lambda_{1}\bxi_{1}(s)-\lambda_{2}\bxi_{2}(s))}
\label{barecoulomb}
\end{align}
with corresponding paths $\br_{1}+\lambda_{1}\bxi_{1}(s),\;\br_{2}+\lambda_{2}\bxi_{2}(s)$ becomes a screened effective potential
\begin{align}
\widetilde{V}_\text{screen}(\bk,\bxi_{1},\bxi_{2})=\widetilde{V}_\text{screen}^{\text{exp}}(\bk,\bxi_{1},\bxi_{2})+\widetilde{V}_\text{screen}^{\text{alg}}(\bk,\bxi_{1},\bxi_{2})
\label{screenedcoulomb}
\end{align}
which is the sum of two contributions \cite{BMA2002} \cite{ACM2007} (section 3.2).
The first one
\begin{align}
\widetilde{V}_\text{screen}^{\text{exp}}(\bk,\bxi_{1},\bxi_{2})
=\frac{4\pi e_{1}e_{2}}{k^2+\kappa^2}\int_{0}^1ds
\e^{\moni\bk\cdot(\lambda_{1}\bxi_{1}(s)-\lambda_{2}\bxi_{2}(s))}
\label{Vexp}
\end{align}
has the  Debye-H\"uckel form familiar in the classical theory of screening. The replacement $4\pi/k^2 \rightarrow 4\pi/(k^2+\kappa^2)$, with $\kappa=\lscreen^{-1}$ the inverse screening length, leads to an exponentially fast decay on scale $\lscreen$ as $|\br_{1}-\br_{2}|\to\infty$.
In the limit of high atomic dilution, the second contribution takes the form 
\begin{align}
&\widetilde{V}_\text{screen}^{\text{alg}}(\bk,\bxi_{1},\bxi_{2})=-
\frac{4\pi e_{1}e_{2}}{k^2}\int_{0}^1ds_{1}\int_{0}^1ds_{2}\[\delta(s_{1}-s_{2})-1\]\e^{\moni\bk\cdot(\lambda_{1}\bxi_{1}(s)-\lambda_{2}\bxi_{2}(s))}
\nonumber\\
&\sim \frac{4\pi e_{1}e_{2}}{k^2}\int_{0}^1ds_{1}\int_{0}^1ds_{2}
\[\delta(s_{1}-s_{2})-1\](\bk\cdot\lambda_{1}\bxi_{1}(s_{1}))(\bk\cdot\lambda_{2}\bxi_{2}(s_{2})), \quad \bk \to 0
\label{Valg}
\end{align}
The spatial decay $\sim |\br_{1}-\br_{2}|^{-3}$ of the corresponding potential  is dipolar and has a pure quantum origin. It represents 
the interaction of  the two fluctuating dipoles $\lambda_{1}\bxi_{1}(s_{1})$ and $\lambda_{2}\bxi_{2}(s_{2})$ generated by the intrinsic fluctuations of quantum charge positions. Here the screening is non exponential, causing only a reduction of the bare Coulomb decay $\sim |\br_{1}-\br_{2}|^{-1}$ to the dipolar one $\sim |\br_{1}-\br_{2}|^{-3}$. This algebraically decaying term disappears
whenever one or both charges are classical. A more thorough discussion of quantum screening effects can be found in 
\cite{Cornu1996I, Cornu1996II, BrydgesMartin1999,Alastuey1999}.
When $|\br_{1}-\br_{2}|\gg \lscreen$ one can disregard the exponentially small contribution of $\widetilde{V}_\text{screen}^{\text{exp}}$ but one must of course keep the long range part $\widetilde{V}_\text{screen}^{\text{alg}}$.

In our model, when $X\gg\lscreen$, this entails replacing the bare Coulomb interaction of the electron with its image by $\widetilde{V}_\text{screen}^{\text{alg}}$ and thus, according to (\ref{Valg}), replacing 
(\ref{PhicFouriertransformb}) by
\begin{align}
&\widetilde{\Phi}_\text{screen}^{\text{alg}}(\bk)=\nonumber\\
&\frac{2\pi e^2}{k^2} \int D_\text{at}^\text{b}[\balpha]\int_0^1 ds_{1}\int_0^1 ds_{2}\[\delta(s_{1}-s_{2})-1\]\,(\bk\cdot\balpha(s_{1})(\bk\cdot\balpha^*(s_{2}))+\cO\(k^2\)\nonumber\\
&=\widetilde{\Phi}_\text{im}(\bk)-\frac{2\pi e^2}{k^2} \int D_\text{at}^\text{b}[\balpha]\int_0^1 ds_{1}\int_0^1 ds_{2}\,(\bk\cdot\balpha(s_{1})(\bk\cdot\balpha^*(s_{2}))+\cO\(k^2\)
\label{PhicFouriertransformscreened}
\end{align}
Using rotational invariance and neglecting fine structure constant effects in the atomic measure as before, one recognizes that the second term is identical to $-2\pi e^2(k_x^2/k^2)B_{\text{class}}(k)$ (see (\ref{Bclass}) and (\ref{classical1})).
Thus the net potential (\ref{netpotential}) is modified to 
\begin{align}
\Phi_\text{screen}^{[0]}(X)\sim 4\pi e^2\int \frac{d\bk}{(2\pi)^3} e^{-\moni k_x 2X} 
\frac{k_x^2}{k^2}\[B(k)-B_{\text{class}}(k)\]=\Phi^{[0]}(X)-\Phi_\text{class}(X)
\label{netpotentialscreened}
\end{align}
when $ X\gg \lscreen$ and remains unchanged when $ X\ll \lscreen$. We summarize below the long distance behaviour of the atom-wall potential according to the value of $\lscreen$ compared to the other lengths $\lph$ and $\lat$. 

\begin{itemize}
\item 
$\lat\ll\lph\ll\lscreen$. If $X\ll \lscreen$ the behaviour of $\Phi^{[0]}(X)$ is that given in subsections 4.3.1, 4.3.2 and 4.1.3. If $ X\gg\lscreen$, we see from (\ref{netpotentialscreened}) and (\ref{classical2}) that there is an exact compensation , $\Phi^{[0]}(X)\sim 0$. Charges and thermalized photons conspire to cancel the leading order term $\sim X^{-3}$.\\
\item
$\lat\ll\lscreen\ll\lph$. If $X\ll \lscreen$, $\Phi^{[0]}(X)$ behaves as in subsections 4.3.1 and 4.3.2. However one has 
\begin{align}
\Phi^{[0]}(X)\sim\Phi_\text{CP}(X)-\Phi_\text{class}(X),\quad      \lscreen\ll X\ll\lph
\label{A}
\end{align}
and $\Phi^{[0]}(X)\sim 0$ when $X\gg\lph$.\\
\item
$\lscreen\ll\lat\ll\lph$.  If $X\ll \lscreen$, $\Phi^{[0]}(X)$ behaves as in subsection 4.3.1,  but
\begin{align}
&\Phi^{[0]}(X)\sim\phi_\text{vdW}(X)-\Phi_\text{class}(X),\quad      \lscreen\ll X\ll\lat\nonumber\\
&\Phi^{[0]}(X)\sim\Phi_\text{CP}(X)-\Phi_\text{class}(X),\quad      \lat\ll X\ll\lph
\label{B}
\end{align}
and $\Phi^{[0]}(X)\sim 0$ when $X\gg\lph$.
\end{itemize}
The various formulae are summarized in Fig.2a Fig.2b and Fig.2c.
where $\Phi^{(2)}$ denotes the diamagnetic contribution to the decay mentioned in subsection 4.1. 
\begin{figure}
\begin{center}
\subfigure[]{\includegraphics[width= \textwidth]{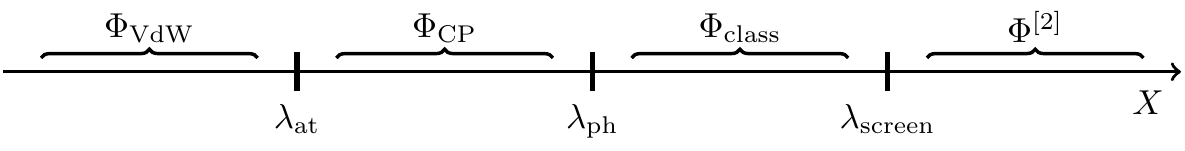}}
\\
\subfigure[]{\includegraphics[width= \textwidth]{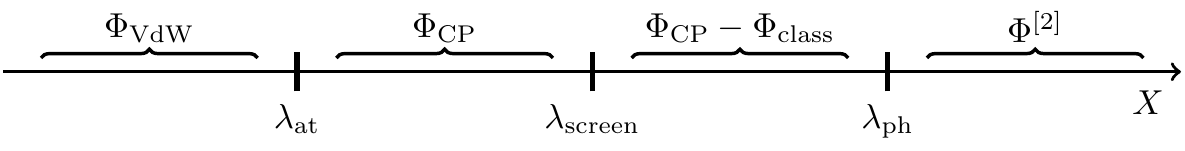}}
\\
\subfigure[]{\includegraphics[width= \textwidth]{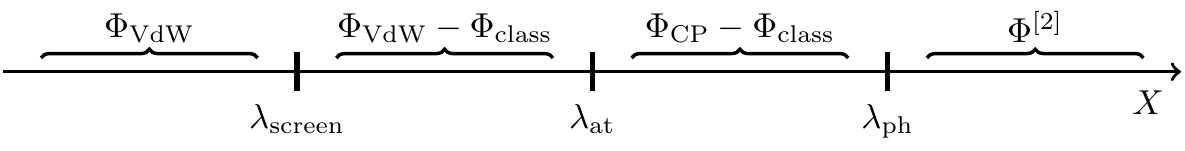}}
\end{center}
\caption{Low-temperature atom-wall potential when $\lscreen$ is of the same order as some characteristic length scale of the microscopic model.}
\end{figure}
The thermal corrections in (\ref{A}) and (\ref{B}), linear in $T$ (see (\ref{classical2})), are analogous to those found for van der Waals
potentials at finite temperature in \cite{ACM2007}.
They account for free charge screening, and because of the negative sign, leading to a weakening of the standard dipole (\ref{Phicas}) and Casimir-Polder (\ref{retarded2}) interactions.

\section{Concluding remarks}
In this paper we have developed a functional integral method
to analyze dispersive forces at a microscopic level taking retardation effects into account. This formalism enables one to
extract the asymptotic form of the force without recourse
to perturbation with respect to the electromagnetic field coupling.
 Miyao and Spohn \cite{MiyaoSpohn2009} have applied  the same method
to give a non-perturbative derivation of van der Waals forces in the atomic ground states.
Here we have considered the atom-wall forces at non zero (but low) temperature and shown in a first stage how to recover van der Waals, Casimir-Polder and classical Lifshitz forces in the dipole approximation.
We have pointed out a consequence of screening that is not obtained in the usual applications of the Lifshitz theory.
If the Coulomb potential between the atom and the wall mirror charges is screened by ionized electrons or possibly other types of mobile charges, we find a temperature correction to the van der Waals or Casimir-Polder force which is linear in $T$. This correction originates from the fact that the screening of quantum charges is not exponential, but algebraic, and thus participates in building up
 the long tail of dispersive forces. Algebraic screening entails non analytic terms in the small wavenumber expansion of the dielectric function $\epsilon(\bk,\omega)$  \cite{CornuMartin1991}. 
 Therefore applications of the Lifshitz theory that use simple analytic forms of dielectric functions to describe the different media do not predict the modifications of dispersive forces due to algebraic screening. 
 
 Moreover the spontaneous emission of a photon by an excited state with energy $E_i$ gives rise to  $\cos[2 X \hbar c/(E_i-E_0)]$ oscillations with a damping $1/X$ factor in the large-distance behavior of the atom-wall interaction. Such  a decay comes up in our treatment (see the remark after \eqref{G}) as well as in 
 the study of an atom prepared in an excited state \cite{MeschedeEtAl1990} or in the calculations for a simplified model for an atom with only two energy levels \cite{MendesFarina2007}, as already pointed out  in the introduction. This  tail, which  has an exponential thermal weight 
 $\exp\[-(E_1-E_0)/\kB T\]$, has been disregarded in the low-temperature limit considered in the present paper. However spontaneous emission effects   should be retained in  thermal corrections whenever such exponential contributions can no more be neglected at higher temperature.

\par\medskip
{\large\textbf{Acknowledgments}}

\par\medskip
F. Cornu acknowledges fruitful discussions with Michel Bauer about stochastic integrals. Ph. Martin thanks the ESF Research Network CASIMIR for providing the opportunity for useful discussions on fluctuation-induced forces and KITP for its kind hospitality. This research was  supported in part by the National Science Foundation under Grant No. PHY05-51164.


\end{document}